\documentclass[review]{elsarticle}

\usepackage{lineno,hyperref}
\usepackage{graphicx}
\usepackage{amsmath}
\usepackage{amssymb}
\modulolinenumbers[5]

\journal{ }

\newcommand{\yb}{\boldsymbol{y}}

\newcommand{\kb}{\boldsymbol{k}}
\newcommand{\vb}{\boldsymbol{v}}
\newcommand{\ab}{\boldsymbol{\alpha}}

\begin{document}

\begin{frontmatter}

\title{Computational framework for real-time diagnostics and prognostics of aircraft actuation systems}


\author[mymainaddress]{Pier Carlo Berri\corref{mycorrespondingauthor}\fnref{fnote1}}
\cortext[mycorrespondingauthor]{Corresponding author}
\ead{pier.berri@polito.it}
\fntext[fnote1]{PhD Student, Department of Mechanical and Aerospace Engineering (DIMEAS), Politecnico di Torino}

\author[mymainaddress,fnote2]{Matteo D.L. Dalla Vedova}
\fntext[fnote2]{Assistant Professor, Department of Mechanical and Aerospace Engineering (DIMEAS), Politecnico di Torino}

\author[mymainaddress,mysecondaryaddress,fnote3]{Laura Mainini}
\fntext[fnote3]{Visiting Professor, Department of Mechanical and Aerospace Engineering (DIMEAS), Politecnico di Torino. Research Affiliate, Massachusetts Institute of Technology, Cambridge, MA 02139, USA}

\address[mymainaddress]{Politecnico di Torino, c.so Duca degli Abruzzi 24, 10129 Turin (IT)}
\address[mysecondaryaddress]{Massachusetts Institute of Technology, Cambridge, MA 02139, USA}

\begin{abstract}

Prognostics and Health Management (PHM) are emerging approaches to product life cycle that will maintain system safety and improve reliability, while reducing operating and maintenance costs. This is particularly relevant for aerospace systems, where high levels of integrity and high performances are required at the same time.
We propose a novel strategy for the nearly real-time Fault Detection and Identification (FDI) of a dynamical assembly, and for the estimation of Remaining Useful Life (RUL) of the system. 
The availability of a timely estimate of the health status of the system will allow for an informed adaptive planning of maintenance and a dynamical reconfiguration of the mission profile, reducing operating costs and improving reliability.
This work addresses the three phases of the prognostic flow -- namely (1) signal acquisition, (2) Fault Detection and Identification, and (3) Remaining Useful Life estimation -- and introduces a computationally efficient procedure suitable for real-time, on-board execution.
To achieve this goal, we propose to combine information from physical models of different fidelity with machine learning techniques to obtain efficient representations (surrogate models) suitable for nearly real-time applications.
Additionally, we propose an importance sampling strategy and a novel approach to model damage propagation for dynamical systems.
The methodology is assessed for the FDI and RUL estimation of an aircraft electromechanical actuator (EMA) for secondary flight controls. The results show that the proposed method allows for a high precision in the evaluation of the system RUL, while outperforming common model-based techniques in terms of computational time.

\end{abstract}


\end{frontmatter}


\section{Introduction}

The steadily increasing complexity of aircraft systems results in large amount of heterogeneous components to integrate.
Each component is characterized by its own set of failure modes, which can interact with those of the other components, increasing the overall system failure rate and making the fault identification and isolation process difficult and time expensive. This can eventually lead to worsen the reliability and availability characteristics of the vehicle.
The traditional approach to system life-cycle management is based on scheduling maintenance interventions \textit{a priori}: components are replaced at the end of their design life, regardless their actual health status
\cite{electronicsReliability, mechanicsReliability, Venkataraman2017}.
This strategy leads to high maintenance costs and cannot guarantee that no failure will occur before the predicted end of life, for example as the result of an undetected manufacturing defect; to reduce risk on safety-related equipment, critical components are redounded \cite{Garmendia2015, CuiWang2015}, increasing weight and further reducing basic reliability.
Conversely, latest approaches like Condition Based Maintenance (CBM) \cite{Pipe_2008,Lv_2015,Zerhouni_2016} and Integrated Vehicle Health Management (IVHM) \cite{Sudolsky_2007,Dunsdon_2008,Jennions_2011} aim to account for advances in Prognostics and Health Management (PHM) disciplines, in order to better manage the maintenance schedule, reducing costs and increasing mission reliability
\cite{Sutharssan_2015,benedettini,williams,Ladj_2016,Yan_2014}.
PHM relies on continuous monitoring of the actual health status of components, to adaptively estimate the system Remaining Useful Life (RUL)
\cite{Rigamonti_2018,Polverino_2016,JIHIN_2017}.
The benefits promised by CBM and IVHM motivate the great interest in enabling next generation systems and vehicles to autonomously detect damages and faults at their early stage, and predict the associated RUL during operations. This capability would allow to replace components only when really needed, avoid disposing systems that are still healthy, and even recalibrate systems operational envelope to guarantee a longer and safer system life.

Common approaches to PHM leverage either model-based strategies (i.e. relying on physics-based representations of the monitored system \cite{Ksenia2016, BattipedeRomeo, Shi2018, JiaHaimin2019}) or data-driven methods \cite{Bailey, MDVdeFano, LEI2020106587}.
A review of model-based condition monitoring strategies to enable system prognostics is provided by 
Tinga and Loendersloot \cite{Tinga2019}.
In \cite{ENGELBERTH2018} a structured residual between the system response and a digital twin is compared to a threshold in order to detect faults of industrial equipment.
In \cite{Cocconcelli2018} faults are detected online with a data-driven algorithm, and later identified offline employing a model-based strategy.
Henry et al. \cite{Henry2019} propose to compare attitude command and measurement of the inertial platform to determine failures in the attitude control system of a spacecraft.
Huang et al. \cite{Huang2017} and Zhao et al. \cite{ZHAO2019213} provide reviews of data-driven approaches to prognostics leveraging statistical methods and deep learning.
In \cite{Qian2019} an Extreme Learning Machine (ELM) is employed for fault detection of wind turbines, while in \cite{BEKTAS2019383} feedforward networks are used for similarity-based prognostics.
Autoregressive integrated moving average (ARIMA) is applied to the RUL prediction of milling machine cutting tools in \cite{Liu2019}.
Model-based techniques usually require large computational resources, and cannot be executed in real-time by on-board hardware. Data-driven methods, conversely, need large datasets for training, which are usually not available from field: as an example, few field data is available regarding the system-level effects of uncommon but critical failure modes.

This paper proposes a computational framework for a nearly real-time estimation of the Remaining Useful Life for dynamical assemblies from measurements available from installed feedback or diagnostic sensors. Those can be of heterogeneous nature: for example, current flowing inside an electric circuit, position and speed of an actuator, pressure and temperature of hydraulic fluid at given locations of the system. 
The methodology combines an optimal signal compression strategy with reduced order modeling and machine learning techniques; this allows to obtain a computationally efficient map from the measured signals to the RUL, and to reduce the storage and processing power required for on-board, time and resource-constrained computations.
Our strategy learns surrogate models of the system offline: online, these surrogate models are employed to speed up the computational burden associated with the determination of the current system health condition and with the estimation of the RUL. Additionally, offline we determine the location of a set of informative components of the monitored signals to store and process; those are employed online to reduce the dimensionality of the problem.

As an application of our methodology, we consider the case of actuators for aircraft flight control systems (FCSs). FCSs are critical aircraft systems because a failure can lead to the impossibility to control the vehicle, with catastrophic consequences. Hence, health monitoring for FCSs has great potential to bring significant improvements in terms of mission reliability, operating costs, aircraft performance, and eventually relax requirements on system redundancies. The problem is inherently challenging: the models of FCS equipment need to combine different disciplines, as mechanical, aerodynamic, structural, hydraulic, and electrical/electronic subsystems operate together to achieve the required performances. The number of possible failure modes is high, and so the dimensionality of the FDI problem. Additionally, different faults may result in similar effects on the system behavior, or particular operating conditions may be misidentified as faults.
All these aspects make this application an interesting demonstration case for the proposed strategy, as they highlight the shortcomings of current approaches.

In this manuscript, Section \ref{PHM} introduces the general formulation of the problem, Section \ref{meth} details the methodology we propose for the prognostic analysis, Section \ref{application} presents the demonstration problem discussed in this paper and the associated physical models, and Section \ref{app_res} presents the results of our investigations.

\section{Prognostics and Health Management (PHM): problem formulation} \label{PHM}

\begin{figure}[b!]
\centering
\includegraphics[width=3.5cm]{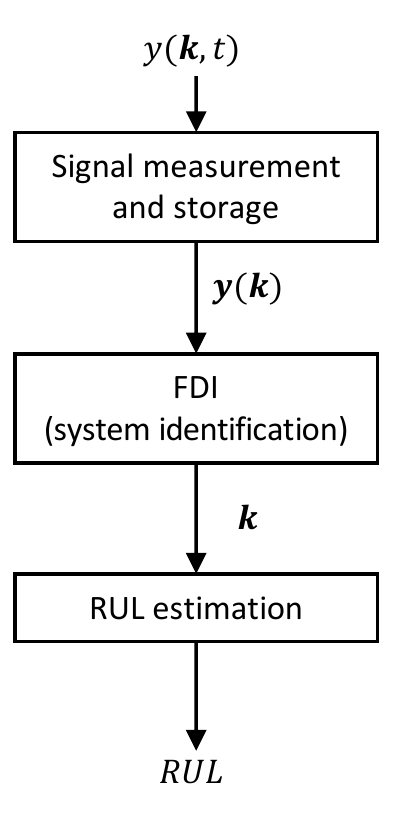} 
\caption{Schematic representation of the ideal RUL estimation flow}
\label{strategy}
\end{figure}

The common prognostics flow includes three steps, namely signal measurement and storage, Fault Detection and Identification (FDI) and estimation of Remaining Useful Life (RUL), as depicted in Figure \ref{strategy}.
In the signal measurement and storage phase, an output signal is measured from the system with a standard acquisition frequency. The signal, sensitive to the system condition, is an indicator of the health status of the components, and can be used to inform the subsequent FDI phase.
In the FDI phase, the system output is processed to identify the early signs of damage and wear.
Eventually, in the RUL estimation phase, the identified health condition is used to inform an estimate of the actual remaining useful life of the system.

In a traditional PHM process, the signal measurement and storage phase is the only one performed in real-time. It consists in acquiring data $y(\kb, t)$ from the available sensors installed on the system. The acquisition frequency is usually fixed, and results from a compromise between hardware capabilities and retained information. The result of the acquisition phase is a vector $y(\kb)$, dependent on the health condition of the system $\kb$. Large amounts of data can be easily produced in this phase, which can be cumbersome to store and to deal with in the following phases of the PHM process: this motivates why the FDI and RUL estimation tasks are usually performed offline. We address this issue aiming to compress the useful information in order to reduce the dimensionality of the FDI problem.

The subsequent FDI phase estimates the current health condition of the monitored system by processing the signals acquired and compressed in the previous step. Common approaches to FDI rely on the use of models, reliable representations of the physical systems and emulators of their dynamic behavior: a system output signal, sensitive to the damage condition, is measured and compared to the output signal computed with a numerical model.
In \cite{Freeman2013}, a physics-based model of aircraft flight dynamics is evaluated to compute the residual between the response of the physical system and its digital twin; then a statistical anomaly detection algorithm analyzes this residual to identify faults of the aileron actuation.
A similar approach is proposed in \cite{Venkataraman01} to determine anomalous behavior of the flight control actuator of a UAV; the strategy analyzes the effects of the failure at aircraft level: as a result, incipient faults are not detectable.
In \cite{Meng_2018}, a dynamical observer leveraging a Kalman Filter is employed for the model-based condition monitoring of wind turbines.
Hence, the FDI problem is a system identification problem whose solution (the current fault condition $\kb^{c}$) is the one that minimizes (ideally vanishes) the discrepancies between the measured signal $\yb$ and the simulated one $\yb^\textrm{model}(\kb)$:

\begin{equation}
\label{FDI_formulation}
\kb^{c} = \textrm{arg}\min_{\kb} (err_y(\yb, \yb^\textrm{model}(\kb))
\end{equation}

\noindent where, in the most general case, the error function $err_y(\yb, \yb^\textrm{model}(\kb)$ is a monotonically increasing function of $\Vert\yb-\yb^\textrm{model}(\kb)\Vert$; the particular norm to be used may vary, and usually is chosen depending on the peculiar characteristics of the measured signals.
If a purely model-based technique is employed, the computation of $\yb^\textrm{model}(\kb)$ is usually expensive. The need to evaluate the error function iteratively within an optimization algorithm leads to computational times incompatible with real-time execution; additionally, the definition of a proper error function may be challenging. Conversely, data-driven strategies are faster, but require large datasets for training, as highlighted by Booyse et al. \cite{BOOYSE2020106612}. Such amounts of field data are often unavailable, especially during the design and validation of equipment, since their collection can only be carried out with several thousands of hours of operation of such equipment.
The health condition $\kb$ determined with FDI is employed for the estimation of Remaining Useful Life.

The RUL of a system is the remaining time until the system will no more be able to meet its functional or performance requirements, that is, the time when the system will not be able to perform its function either at all or within the design performance parameters \cite{Vachtsevanos_2006, Isermann_2011}. 
This definition can be formalized as:

\begin{equation}
\label{RUL:definition}
\begin{aligned}
& \text{RUL} = \max(t) \\
& \text{s.t.} \quad \phi_a(\kb(t)) = \textrm{``healthy''}
\end{aligned}
\end{equation}

\noindent where $\phi_a(\kb)$ is an assessment function. $\phi_a(\kb)$ is a binary valued function assuming the possible values $\textrm{``healthy''}$ or $\textrm{``faulty''}$ that determines whether the fault vector $\kb$ corresponds to a healthy system or not (i.e. whether the system is still compliant to its functional and performance requirements).

With the traditional approach to life cycle management, the system useful life is computed \textit{a priori} in the design phase, solely from the probabilistic combination of components failure rate. This strategy does not account for the real evolution of the components health status, and then produces estimates affected by a very large uncertainty interval \cite{electronicsReliability, mechanicsReliability, Venkataraman2017, Garmendia2015}.
Popular approaches to RUL estimations aim at obtaining a more precise estimate of the system life either by extrapolating the current fault propagation rate \cite{Yongxiang2016}, or by employing a model of damage growth until the damage condition reaches a threshold.
In \cite{Al_Dahidi_2016}, a statistical approach combines a semi-markov model and the Maximum Likelihood Estimation (MLE) method to infer a degradation model for the equipment.
Nascimento and Viana \cite{NASCIMENTO_2019} discuss the use of recurrent neural networks merging physics-informed and data-driven knowledge to model the time evolution of structural fatigue.
Jacazio et al. \cite{JacazioRUL, DeMartin2020} propose to employ particle filtering to estimate the system RUL; in \cite{LI2019} particle filtering is combined with Canonical Variate Analysis (CVA) and Exponentially Weighted Moving Average (EWMA) in order to determine the RUL of rotating equipment.
However, these methods often require a significant computational effort, or may be highly influenced by the effect of uncertainty in the estimation of the fault condition.
Additionally, the definition of a proper critical failure threshold may be difficult: usually individual thresholds are set for each considered failure mode, not accounting for the combined effect of multiple faults.
These can affect the system performance in a different way than the linear superposition of the effects of individual faults. As a result, a more general and comprehensive definition of a critical failure level may be needed.

Active research in PHM aims to enable early estimate the RUL, much in advance to the actual failure event, in order to allocate time for the optimal planning of maintenance strategies and 
for the logistics of fleet management.
This motivates the interest for advanced FDI procedures to detect incipient faults at their early stages, before the system-level performances of the equipment start becoming significantly and adversely affected.
To capture incipient faults, we specifically developed an importance sampling strategy (Section \ref{MLH}) for the computation of the dataset needed for training the machine learning tools. The proposed technique is intended to get denser sampling for small faults, where most useful information is expected. Nevertheless, the choice of an adequate sampling procedure is problem dependent, and other sampling strategies can outperform the proposed one on different applications of the same methodology.

Both FDI and RUL estimation tasks imply the execution of a system emulator: usually this model is associated with an expensive computational effort.
Therefore, most existing model-based \cite{Ksenia2016, BattipedeRomeo, Shi2018} and data-driven \cite{Bailey, Naderi2017, MDVdeFano} strategies are not suitable for real-time execution.
Specifically, we wish to perform the FDI and RUL estimation tasks on-board, which requires to meet the hardware resources limitations to achieve a nearly real-time process. Therefore specific strategies are needed to achieve such computational efficiency and to meet these constraints.

\section{Methodology} \label{meth}

Our methodology proposes specific combinations of machine learning techniques to address each of the three phases of the flow described in Figure \ref{strategy} in a computationally efficient manner. Specifically, offline we compute surrogate models that are employed to speed up the online computations.

For the first phase of signal acquisition and storage, we aim to reduce the data required to store a system output signal $y(\kb,t)$. A uniform standard acquisition sampling with a suitable frequency produces a vector $\yb(\kb)$ whose size is impractical for the storage and subsequent processing. For example, monitoring a single electromechanical actuator may imply the acquisition of currents and voltages with frequencies in the order of tenths of kilohertz, resulting in a datarate up to several MB/s.
To address this issue, offline we define an optimal signal compression in two steps: projection based model reduction (Proper Orthogonal Decomposition) and unsupervised machine learning (Self-Organizing Maps) are combined to determine a set of informative components of $\yb(\kb)$ to store and process. Online, the compressed output $\hat{\yb}(\kb)$ is a vector containing only the selected informative components of $\yb(\kb)$.
To improve robustness against measurement uncertainty, $\hat{\yb}(\kb)$ is not fed directly to the subsequent phases. Online we adopt Gappy POD to reconstruct POD coefficients from the compressed representation of the signal $\hat{\yb}$; those are used as input for fault estimation.

The second phase is the Fault Detection and Identification (FDI): this step aims at identifying the health condition of the system, i.e. the specific fault vector $\kb$, with limited computational resources.
Offline, we use supervised machine learning (Multi-Layer Perceptron) to compute a model for the fault condition $\kb$ as a function of the coefficients $\boldsymbol{\alpha}$ of the POD expansion. Online we use the surrogate model learned offline to estimate the fault vector $\kb$ from the reconstructed POD coefficients.

The third phase is the RUL estimation. Here, a simple model of damage propagation is evaluated to compute an estimate of the remaining life of the system. In this phase, an estimator of the computationally expensive assessment function is needed as a stopping criterion for the damage propagation. To meet the constraints in terms of time and available processing power, we propose the use of a binary classifier, specifically a Support Vector Machine (SVM) to replace the complete assessment function. The SVM is trained offline on the reference dataset and employed online to speed up the RUL estimation.

\begin{figure}[b!]
\centering
\includegraphics[width=\textwidth]{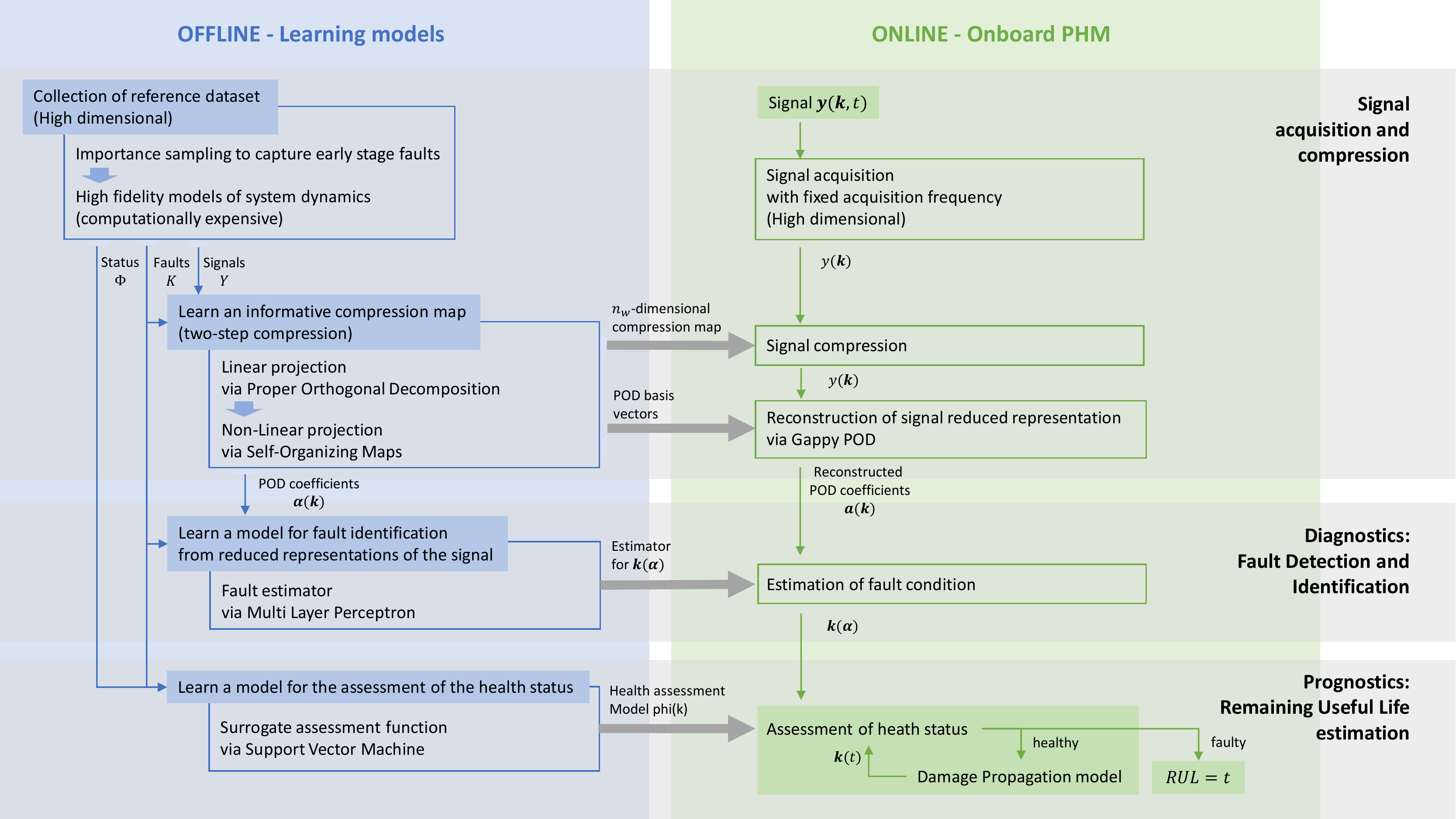}
\caption{Schematic representation of the proposed prognostic flow}
\label{strategy1}
\end{figure}

The strategy is schematically illustrated in Figure \ref{strategy1}: Sections \ref{compression}, \ref{FDI} and \ref{RULest} describe the two steps signal compression, the FDI phase and the RUL estimation procedure, respectively; Section \ref{MLH} describes the collection of the reference dataset used to learn the models (offline).

	\subsection{Importance sampling via particular scaled latin hypercube strategy}  \label{MLH}

Learning the surrogate models requires the collection of a training dataset representative of the system behavior under the expected operating conditions.
It can be collected according to a variety of sampling strategies, depending on the specific problem at hand. For the application discussed in this paper we propose a particular importance sampling method.
The reference data can be collected from a variety of different sources including historical data, numerical simulations of the systems through evaluations of high fidelity models, or experimental measurements. In this paper we use data from high-fidelity, accurate models of the systems, considered as a ground-truth reference.

The data set in organized into the following quantities of interest:

\begin{itemize}
	\item {\sc Fault conditions matrix} $\textrm{K}$:
	$\textrm{K} = [\kb_1, \kb_2, ..., \kb_{n_s}]^\top$ is a $n_s$-by-$n_k$ matrix containing in its rows the $n_s$ fault combinations $\kb_i$ collected in the dataset. $\kb_i$ are the $n_k$-dimensional fault vectors that carry the information about the system health condition.
	Each fault vector encodes in its elements a combination of progressive damages of the system. The elements of the fault vectors are, in general, related to physical quantities of different nature, such as the friction coefficient between two sliding surfaces of a mechanism, the mechanical play of a transmission, or the resistance of an electric circuit; to avoid the effect of different scales and inhomogeneous measurement units, we chose to normalize those quantities, in order to bound the elements of $\kb_i$ between 0 and 1.
	
	\item {\sc Measured signals matrix} $\textrm{Y}$:
	$\textrm{Y} = [\yb_1, \yb_2, ..., \yb_{n_s}]$ is a $n_e$-by-$n_s$ matrix containing in its columns the output vectors $\yb(\kb_i)$ of the system. 
	$\yb(\kb_i)$ are the output signals of the system for each sampled fault combination $\kb_i$, expressed in the form of $n_e$-dimensional vectors by capturing them at a fixed acquisition frequency.
	
	\item {\sc Assessment function matrix} $\Phi$: 
$\Phi = [\phi_{a,1}, \phi_{a,2}, ..., \phi_{a,n_s}]$ is a $1$-by-$n_s$ matrix containing the values of the assessment function corresponding to each fault combination.
$\phi_{a,i} = \phi_a(\kb_i)$ is the value of the assessment function for the fault vector $\kb_i$.
	
\end{itemize}

The particular sampling strategy for collecting the dataset is problem dependent. The application presented in this paper requires to detect the early signs of a system damage with high accuracy in order to determine the Remaining Useful Life in advance enough to plan corrective actions. For this reason, the training dataset shall be denser of health conditions close to the nominal one, that is, when either no faults are present or faults are small and do not have a significant effect on the system performances.
In this context, we implemented a form of importance sampling strategy through a scaled latin hypercube \cite{MORRIS1995381, Keane_2008, Forrester_2006, Viana_2009}.
This technique is meant to increase the density of sampling points near the nominal condition; this allows to collect more informative samples to capture small and incipient faults.

For example, the assessment function is expected to assign $\textrm{``healthy''}$ labels to fault conditions $\kb$ near the nominal condition $\kb_0$, and $\textrm{``faulty''}$ labels to fault conditions far from the nominal one.
Being the number of parameters large, a uniform distribution of points in the domain of $\kb$ would result in a small fraction of points near the nominal condition (i.e. $\textrm{``healthy''}$ fault combinations), and the surrogate assessment function would be difficult to train.
To give an instance, given the eight-dimensional unit hypercube, the points whose distance from the origin is smaller than 1 account for less than 1.6\% of the total volume.

To have a significant fraction of sampling points associated to a healthy condition, we proceed with two steps. First, we implement a standard Latin Hypercube sampling to obtain a $n_k$-by-$n_t$ matrix $\textrm{J} = [\boldsymbol{j}_1,...,\boldsymbol{j}_{n_k}]$. Each row of the matrix encodes a sampling point of the training set, and each column is related to one fault parameter.
Then, the points are rescaled to produce a uniform distribution in the distance from the origin, measured with a $L_\infty$ metric.
Assuming that the nominal condition is in the origin, we do not lose of generality, since we can define $\hat{\kb} = \kb - \kb_0$. The matrix $\textrm{K} = [\kb_1,...,\kb_{n_k}]$ is computed from $\textrm{J}$ by scaling its elements near the origin:

\begin{equation}
\textrm{K}_{ij} = (\textrm{J}_{ij})^{n_k}
\end{equation}

Then, the rows of $\textrm{J}$ characterized by $L_\infty (\textrm{J}_{i,:}) \leq L$ are contained in an $n_k$-dimensional hypercube of side $0 \leq L \leq 1$; being the probability distribution of $\textrm{J}$ uniform, their number is approximately:

\begin{equation}
n_t L^{n_k}
\end{equation}

Those points are mapped to points of $\textrm{K}$ contained in $0 \leq L^{n_k} \leq 1$, then $L_\infty (\textrm{K}_{i,:}) \leq L^{n_k}$. Hence, the number of points of $\textrm{K}$ such that $\Vert\textrm{K}_{i,:}\Vert_\infty \leq a$ is proportional to $a$ for any $a \in \mathbb{R}$, resulting in a uniform probability distribution for $\Vert\textrm{K}_{i,:}\Vert_\infty$.
This sampling strategy is employed, in combination with the physics-based models described in Section \ref{EMAmodels}, to obtain the reference data of the matrices $\textrm{K}$, $\textrm{Y}$ and $\Phi$.

\subsection{Signal Acquisition and Compression} \label{compression}
	
The first obstacle that results in a computationally expensive process is the high dimensionality of $\yb$.
Any output signal measured from the system (i.e. currents and voltages of an electrical machine, hydraulic pressures, accelerations etc.) with a uniform sampling is a vector composed by $n_e=\Delta t f_s$ elements, where $f_s$ is the acquisition frequency and $\Delta t$ is the observation time. In most applications, to capture the required amount of information, $f_s$ needs to be in the order of tenths of kilohertz and $n_e$ can easily be in the order of several thousands. This requires large storage capabilities and processing power for the subsequent Fault Detection and Identification phase. For instance, the use of common least squares methods for the subsequent FDI phase involves the QR factorization of an $n_e$-by-$n_k$ matrix, where $n_k$ is the number of fault parameters, which computational cost ($\mathcal{O}(n_e^3)$) makes the online/on-board execution impractical for the large $n_e$ of common output signals $\yb$.

To overcome this problem, we use a particular strategy for the optimal selection of a small number of the sampling points to retain, store and process. A first approach of this kind was introduced by Mainini and Willcox \cite{Mainini_2017}, where Proper Orthogonal Decomposition and Self-Organizing Maps are combined for the optimal placement of sensors for on-board assessment of structural capabilites. In this work a similar approach is adopted to reduce the computational burden associated to the FDI task by reducing the problem dimensionality to $n_w \ll n_e$.

The selection of the signal points to process online is computed offline through a two-step procedure to learn a compression map: it combines low order representations of high dimensional data (projection based model reduction) and machine learning techniques (unsupervised machine learning) to identify the most informative instants of time of the measured signal to be stored for the subsequent online Fault Detection and Identification.	

		\subsubsection{Offline: Learn an informative compression map}
		
We aim to determine an informative compression map for the signal $y(\kb, t)$. Only those points of the signal will be stored and processed online, reducing the required computational resources.
The offline signal compression process takes as input a set of measurements from the system.
For this purpose, we use the fault conditions in $\textrm{K}$ and the associated output signals (snapshots) $\yb$ assembled into the columns of the $n_e$-by-$n_s$ measurement matrix $\textrm{Y}$.
Through the proposed two-steps offline compression strategy, we determine the set of informative time-locations for the signal. The compression process is articulated into the two steps leveraging Proper Orthogonal Decomposition (POD) and Self-Organizing Maps (SOMs) respectively.

			\paragraph{Linear projection via Proper Orthogonal Decomposition (POD)} \label{POD}

The first step of compression employs data gathered by simulations or experimental campaigns, with the purpose of obtaining a reduced order representation of the system. This reduced model is computed through Proper Orthogonal Decomposition (POD).
POD
\cite{Hinze, lumleyPOD,  holmes_lumley_berkooz_1996, Kunisch_1999,  Cand_s_2006}
is a projection based reduced order modeling technique commonly employed to obtain low dimensional representations of high dimensional quantities, through the identification of underlying features (in the form of dominant modes). One of the most employed strategy is the method of snapshots \cite{Sirovich}.
Data points are represented in the $n_e$-dimensional space, and the dominant modes are the principal directions along which the points are dispersed. The eigenvalues associated with each mode encode the variance of the data set along that direction.

We apply POD to the measurements matrix $\textrm{Y}$, in order to extract the modes associated with the largest eigenvalues, that explain most of the variance of the dataset.
The POD modes constitute an orthonormal basis for the measured signals collected in $\textrm{Y}$ \cite{Algazi2006, Dur_1998}: it is optimal in the least squares sense and can be computed through Singular Value Decomposition (SVD) of matrix $\textrm{Y}$ to obtain a representation of each training signal as:

\begin{equation}
\label{POD:equation1}
\yb(\kb) = \yb_0 + \sum\limits_{i=1}^{n_s} {\vb_i\alpha_i(\kb)}
\end{equation}

\noindent where $\yb_0$ denotes a reference signal (in our application the system output in nominal conditions), $n_s$ is the number of snapshots, equivalent to the total number of POD modes, $\vb_i$ are the POD modes and $\alpha_i(\kb)$ are the coefficients of the POD expansion.
The eigenvalue $\lambda_i$ associated to each mode $\vb_i$ is a measure of the dispersion of the high dimensional training data along the direction defined by the mode itself: by considering only the first $n_m$ modes of the POD expansion (equation \ref{POD:equation1}), the fraction of retained information is given by the cumulative sum of the eigenvalues $\sum_{i=1}^{n_m} {\lambda_i} / \sum_{i=1}^{n_s} {\lambda_i}$.
The POD modes are ordered according to their associated eigenvalue, so we can truncate the expansion to retain only the first $n_m \ll n_s$ modes and to get a low dimensional representation of the signal. If the cumulative sum of the retained eigenvalues is close to 100\%, the information lost in the compression is accordingly small; additionally, if the training set is statistically representative of the actual system behavior, the same compression can be applied to signals not belonging to the training set.

Through POD we obtain a set of basis vectors $\vb_i$ and the associated coefficients $\alpha_i$ for each column of the training set $\textrm{Y}$. Bases and coefficients of the POD expansion are employed both offline and online in the following steps of our procedure.

			\paragraph{Non-linear projection via Self-Organizing Map (SOM)} \label{SOM}

In the second step of signal compression we use the first $n_m$ POD basis vectors to find a compressed representation of the basis vectors themselves through a Self-Organizing Map (SOM). This compressed representation is identified in the form of a set of $n_w \ll n_e$ highly informative time-locations for storing and processing the signal.

A Self-Organizing Map is a single layer neural network that can be used to identify subsets of similar data through unsupervised competitive learning 
\cite{Kohonen_1998, Somervuo1999, Kohonen2001}.
The $n_w$ neurons of the SOM have representations in the input space as weight vectors whose values are updated during the training. In this case the input space is the $n_m+1$ dimensional parameter space given by the time coordinate $t$ and the $n_m$ modes of the POD (see Section \ref{POD}).
The training set for the SOM is given by the first $n_m$ modes of the POD and the corresponding time coordinate $\boldsymbol{t}$, arranged in an $n_e$-by-$(n_m+1)$ array $\textrm{T}$:

\begin{equation}
\textrm{T} = [ \boldsymbol{t}, \vb_1, ..., \vb_{n_m} ]
\end{equation}

\noindent where $\boldsymbol{t}$ and $\vb_i$ are column vectors of $n_e$ elements. During training, all the points of the training set are presented to the network multiple times (\textit{epochs}) in a different order, to avoid a training bias. For each training point $\boldsymbol{\tau}_i$ (the $i$-th row of $\textrm{T}$), a winner neuron $l$ is the one whose weight vector $\boldsymbol{w}_l$ is the closest to the input point:

\begin{equation}
l = \textrm{arg}\min_j (\Vert{\boldsymbol{\tau}_i-\boldsymbol{w}_j}\Vert)
\end{equation}

\noindent where $\Vert \cdot \Vert$ denotes the $L_2$ norm adopted as similarity metric for the study discussed in this work. The neighbor neurons are activated according to a neighborhood function, defined in the space of the topological representation of neurons, usually decreasing with the distance from the winner neuron and symmetric about the winner neuron \cite{Kohonen_1998, Kohonen2001}.
One of the key characteristics of SOMs is that during training, the weight vectors of the neurons are updated to represent a non-linear projection of the high dimensional training data (the first $n_m$ POD modes) onto a lower dimensional manifold, where prototype vectors encode representative points of the POD modes \cite{Mainini_2017, MAININI201701}.
As a result, once training is complete, the first components of all the $n_w$ weight vectors encode the most informative time-locations $\hat{\boldsymbol{t}}$ for the signal $y$. 

		\subsubsection{Online} \label{compression_online}
		
Online, these specific values in the time coordinate are used to store and process the measured signals.
Those signals are acquired in real-time by sensors installed on the monitored equipment, with a constant frequency high enough to capture the information related to the considered progressive failures.
This results in a continuous data stream from the sensor to the acquisition electronics, with a rate than can reach the order of megabytes per second for a single sensor. Data measured during the observation time $\Delta t$ could be stored in an $n_e$ dimensional vector $\yb$.
However, leveraging the two steps compression computed offline, we can store only the $n_w$ informative components.
The resulting compressed signal is a $n_w$-by-1 vector $\hat{\yb}(\kb)$, with $n_w \ll n_e$, that preserves the useful information regarding the faults affecting the system.
As such, $\hat{\yb}(\kb)$ is used as informative input for the subsequent phase.

In principle, the compressed signal $\hat{\yb}(\kb)$ can be directly processed for the detections and identification of the associated fault condition; however, $\hat{\yb}(\kb)$ carries measurement noise and a random error on the signal would directly affect the identification of faults (FDI). To mitigate the effect of measurement noise, we propose to compute the POD coefficients $\ab(\kb)$ from the compressed signal $\hat{\yb}(\kb)$ via Gappy POD and move the FDI task onto the reduced space identified by the POD in Equation \ref{POD:equation1}.  

Gappy POD is a procedure derived from Proper Orthogonal Decomposition and is commonly used for the recovery of incomplete data 
\cite{Everson_1995, Bui_Thanh_2004, Willcox_2006, Mainini_2015}.
Given $\hat{\yb}$, the reconstructed signal can be obtained as a linear superposition of the first $n_m$ POD modes computed offline  (Equation (\ref{POD:equation1}), Section \ref{POD}).
The expansion coefficients $\alpha_j(\kb)$ are computed to minimize the squared error between the known points of the compressed signal $\hat{\yb}$ and its reconstruction in the $n_w$ informative elements.
The coefficients $\alpha_j$ are reconstructed by solving the linear system:

\begin{equation}
\textrm{G} \boldsymbol{\alpha} = \boldsymbol{f}
\end{equation}

where $\textrm{G} = \hat{v}^\top \hat{v}$ is the Gappy Matrix and $\hat{v} = [\hat{\vb}_1 \ldots \hat{\vb}_{n_m}]$ is a $n_w$-by-$n_m$ matrix whose columns contain the $n_w$ informative elements of the first $n_m$ POD modes. The vector $\boldsymbol{f}$ is the projection of the compressed signal $\hat{\yb}$ along the compressed POD modes $\hat{v}$:

\begin{equation}
\boldsymbol{f} = \hat{v}^\top \hat{\yb}
\end{equation}

An approximation of the uncompressed signal could be recovered as the linear combination of the first $n_m$ modes weighted by the coefficients $\boldsymbol{\alpha}$, as per Equation (\ref{POD:equation1}).
However, for the purpose of this work, we are not interested in the reconstruction of the original signal, but exclusively in recovering the POD coefficients $\ab(\kb)$.
These are employed in the next step to identify an estimate of the fault condition $\kb$.

	\subsection{Fault Detection and Identification (FDI)} \label{FDI}
	
The Fault Detection and Identification (FDI) phase of our strategy aims at identifying the health condition of the system (the specific fault vector $\kb$) from the information of the compressed signal $\hat{\yb}$. This task is a parameter identification problem and is formulated as an optimization problem (Equation \ref{FDI_formulation}). However,
the use of common gradient-based or meta-heuristic optimization algorithms for parameter identification requires the iterative evaluation of system emulators that are frequently expensive. The accuracy level required for a reliable identification of the fault condition demands for the evaluation of models of the dynamical system that are usually too computationally expensive to evaluate online \cite{Berri_2017, BerriMainini18}.

To meet the efficiency requirements of time constrained online evaluations, we employ feedforward Neural Networks to estimate the fault vector $\kb$ from the compressed representation provided by the reconstructed POD coefficients $\ab$.
Specifically, a Multi-Layer Perceptron (MLP) maps the POD coefficients to the fault vector.
The FDI task is split into an offline phase, in which the Neural Network model is trained, and an online phase, in which it is evaluated to estimate $\kb$.

		\subsubsection{Offline: Learn a model for fault identification from reduced representations of the signal} \label{FDI-NN}

In order to estimate the fault vector $\kb$ within a computational time suitable for real-time evaluation, we train offline a neural network to obtain a surrogate model for the fault condition $\kb$ from the low dimensional representations of the measurement provided by $\ab(\kb)$.

\begin{figure}[t!]
\centering
\includegraphics[width=\textwidth]{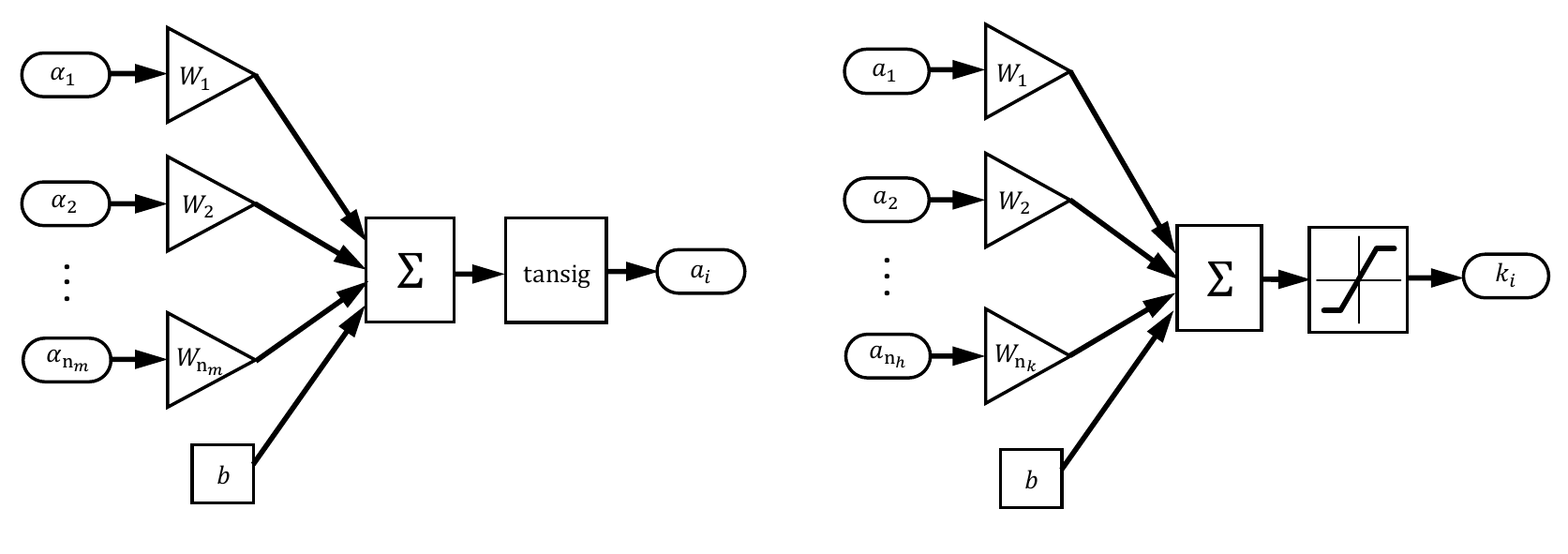} \
\caption{Block diagram of the $i$-th sigmoid neuron of the hidden layer (left) and the $i$-th linear saturated neuron of the output layer (right)}
\label{sigmoid_linear}
\end{figure}

The specific implementation of Multi-Layer Perceptron (MLP) adopted in this paper is characterized by a standard feedforward architecture, with a single hidden layer; more complex machine learning strategies may be tested in future works.
The network receives in input the POD coefficients $\ab(\kb)$ and returns the fault vector $\kb$.
The hidden layer has $n_h$ neurons with sigmoid activation function, while the output layer has $n_k$ neurons with a linear saturated activation function.
The specific choices for the activation functions reflect the physical characteristics of the input and output variables of the problem.

Figure \ref{sigmoid_linear} represents the architecture of a sigmoid and a linear saturated neuron.
The $n_m$ inputs $\boldsymbol{\alpha}$ (column vector) are weighted by the coefficients $\boldsymbol{W}$ (row vector) and summed. Then the weighted sum is fed to a sigmoid function, which returns the output $a$ of the neuron:

\begin{equation}
a = \textrm{tansig}( \boldsymbol{W\alpha} + b )
\end{equation}

\noindent where:

\begin{equation}
\textrm{tansig}(x) = \frac{2}{1+e^{-2x}} - 1
\end{equation}

\noindent and $b$ is a bias constant. The output layer is composed by saturated linear neurons, whose transfer functions are linear saturations:

\begin{equation}
k_i=\begin{cases}
               0, & \text{if $\boldsymbol{Wa}+b<0$}.\\
               \boldsymbol{Wa} + b, & \text{if $0\leq\boldsymbol{Wa}+b\leq1$}.\\
               1, & \text{if $\boldsymbol{Wa}+b>1$}.
    \end{cases}
\end{equation}

\begin{figure}[b!]
\centering
\includegraphics[width=12cm]{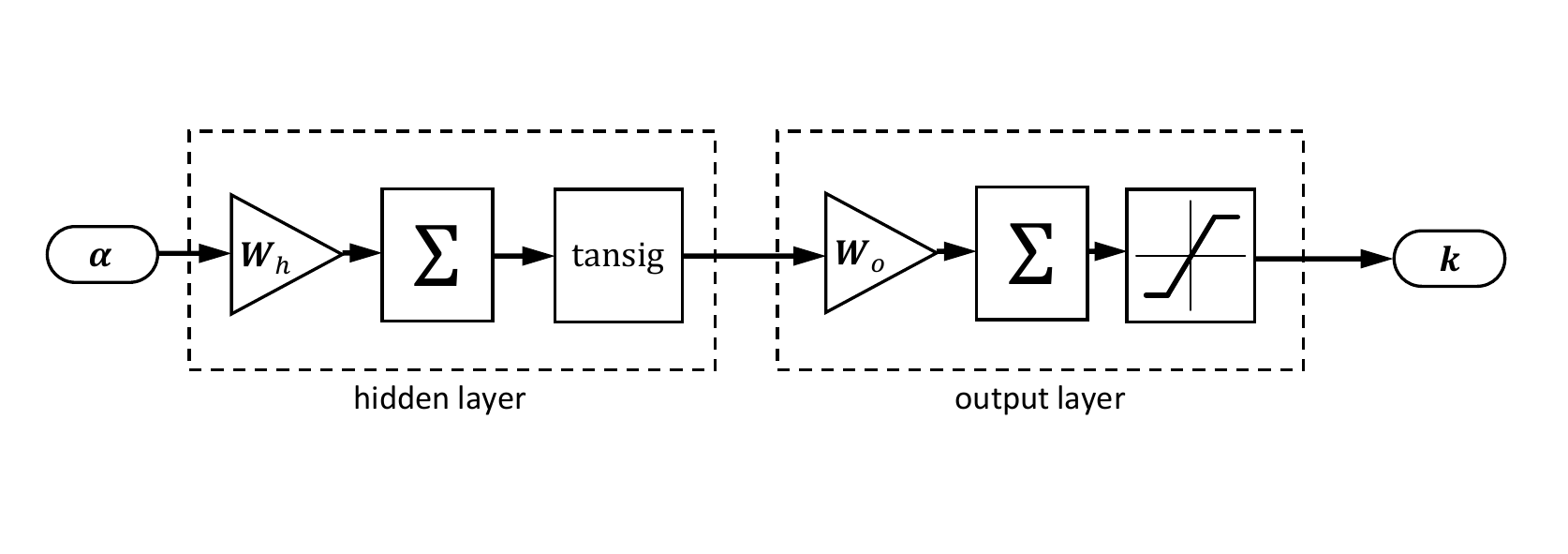} \
\caption{Architecture of the two layer perceptron employed for the FDI task. $W_h$ are the weights of the hidden layer neurons, $W_o$ are those of the output layer}
\label{net}
\end{figure}

The saturation is introduced to account for the bounds of the output fault vector whose components are  bounded between 0 and 1, as defined for our application (Section \ref{EMAmodels}). Figure \ref{net} shows the complete network architecture.
The weights $\boldsymbol{W}$ and the bias $\boldsymbol{b}$ of each neuron are determined during training, in order to tune the network to approximate the expected output for a training data set.
The training dataset is given by the input-output pairs $(\ab_i, \kb_i)$, including the fault vectors $\kb$ collected with the sampling strategy detailed in Section \ref{MLH} and the associated POD coefficients computed as per procedure described in Section \ref{POD}.

During training, a Levenberg-Marquardt backpropagation algorithm 
\cite{Marquardt63, Hagan_1994}
updates the weight and bias variables to minimize a performance function, defined as the mean squared error between the expected and actual output of the network for the training set.
The training is stopped when either the maximum number of epochs is reached or the performance gradient decreases below a threshold.
Once training is complete, we obtain a model to map from $\boldsymbol{\alpha}$ to $\kb$ using the $n_s$ training signals.

		\subsubsection{Online: Estimation of the fault condition}

The input of this phase are the POD coefficients $\ab$ estimated via Gappy POD, as per Section \ref{compression_online}. Those are fed to the MLP model learned offline, in order to estimate the faults $\kb$.
This approach is preferred over the straight adoption of a neural network over the full dimensional dataset  because the compression allows to significantly reduce the computational cost, both in training and in evaluation of the MLP \cite{PBML}.
The output of the FDI process is an estimate of the fault vector $\kb$, to be employed in the subsequent RUL estimation.

	\subsection{Estimate the Remaining Useful Life (RUL)} \label{RULest}

The estimation of the Remaining Useful life is the last phase of the PHM process. We aim to complete it onboard, given the fault condition $\kb$ estimated through the FDI procedure discussed in Section \ref{FDI}.

In this paper we propose a strategy for RUL estimation relying on a damage tolerant approach to system design similar to that adopted for the estimation of fatigue life in aircraft structures.
Leveraging the definition introduced by Equation (\ref{RUL:definition}),
the heath state $\kb$ detected at the mesurement time $t_0 = 0$ is used as an initial value to compute the evolution of the health condition through a specific model for damage propagation. The damage propagation model is in the form of an Ordinary Differential Equation (ODE) whose evaluation provides the rate of damage growth as a function of the current system health and the operating and environmental conditions.
An assessment function $\phi_a(\kb)$ is employed as a stopping criterion for the integration of the ODE model: according to the definition of Equation (\ref{RUL:definition}), it
evaluates the system performances for each value assumed in time by the fault vector, to determine whether that specific fault vector is compatible with the system operation.
When a faulty system is detected, the integration is stopped and the last time step is assumed as the system RUL.

The damage propagation model in the form of ODE may not be computationally expensive since the fault propagation rate is considered to be affected by a limited number of factors (heat dissipation, vibration levels, degradation of surface finish).
Conversely, the evaluation of the assessment function $\phi_a(\kb)$ usually implies multiple executions of the models of system dynamics, which is computationally expensive and unsuited for nearly real-time applications.
To address this limitations we propose to use supervised learning techniques (specifically Support Vector Machines, SVMs), which are trained offline on a reference dataset to obtain surrogate models of the assessment function to employ online.

		\subsubsection{Offline: Learning a model for the assessment of the health status} \label{SVMmodel}

The assessment function $\phi_a(\kb)$ is essentially a binary classifier: it analyzes the behavior of the system in presence of the fault combination $\kb$ and determines whether in this condition the functional and performance requirements are met. This process usually involves the resolution of a dynamical model of the system and requires a high computational effort.
To meet the time constraints for on-board estimation, we adopt a standard implementation of a Support Vector Machine (SVM) as a binary classifier and a surrogate for the assessment function $\phi_a(\kb)$ to run online.

\begin{figure}[b!]
\centering
\includegraphics[width=\textwidth]{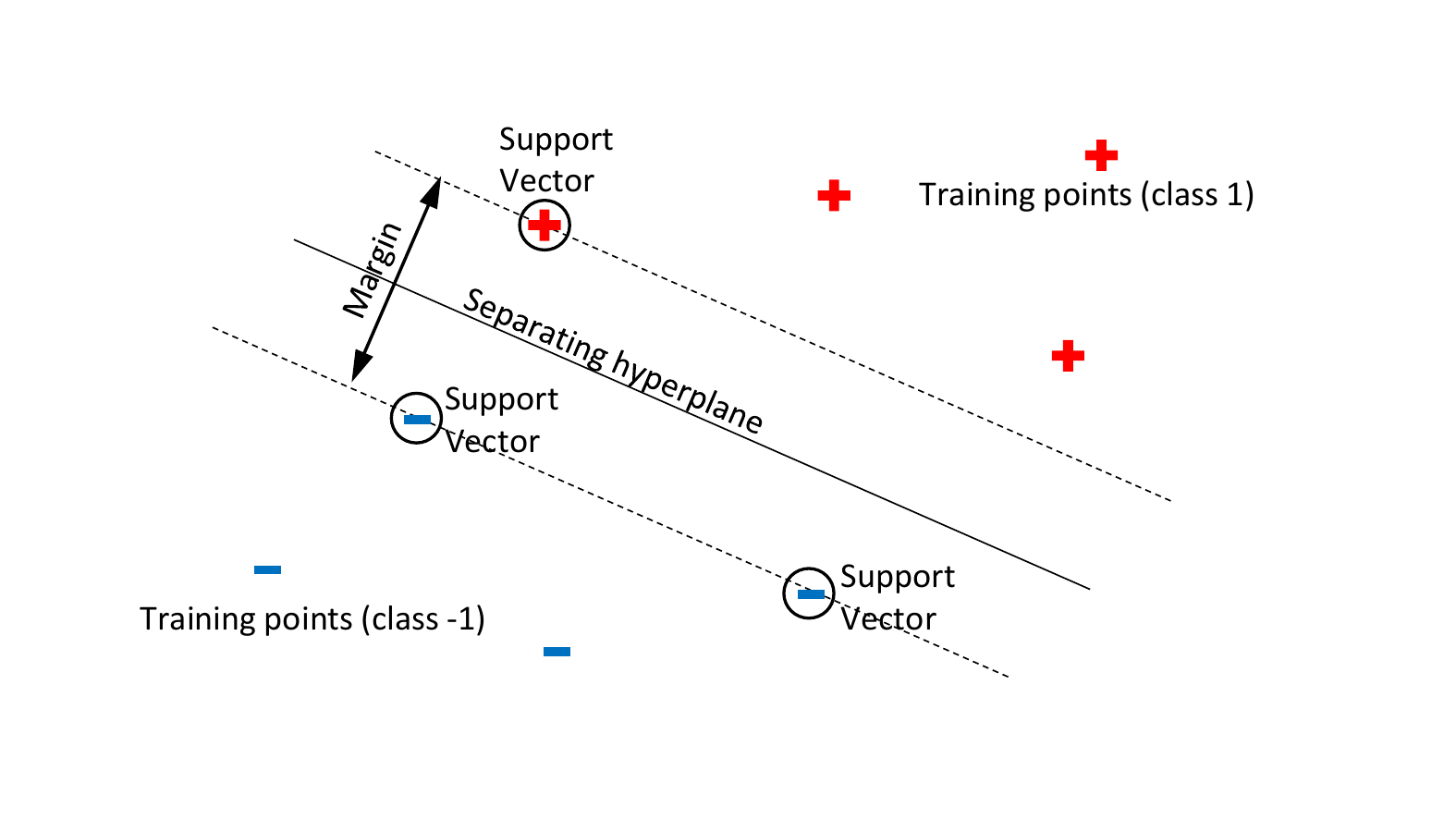} \
\caption{Graphical representation of an SVM classifier}
\label{figure:SVM}
\end{figure}

A Support Vector Machine 
\cite{Cortes1995, Gestel2004BenchmarkingLS, Leng_2018}
is a machine learning paradigm commonly used for data classification and regression.
According to the standard linear formulation, given the training set $\textrm{K}$ of $n_s$ fault conditions $\kb_i \in \mathbb{R}^{n_k}$ and their categories $a_i = \pm 1$, we seek the hyperplane in $\mathbb{R}^{n_k}$ separating the categories $a_i$ (Figure \ref{figure:SVM}):

\begin{equation}
f(\kb) = \kb^\top \boldsymbol{\beta} + b
\end{equation}

\noindent where $\boldsymbol{\beta} \in \mathbb{R}^{n_k}$ and $b \in \mathbb{R}$. The training of the SVM searches for the best hyperplane which divides the classes of $\kb_i$, that is, the hyperplane that produces the largest margin between the classes (see Figure \ref{figure:SVM}); this is equivalent to find $\boldsymbol{\beta}$ and $b$ that minimize $\Vert \boldsymbol{\beta} \Vert$, subject to $a_i f(\kb_i) \geq 1$. The optimization is a quadratic programming problem, and the training algorithm implements a Lagrange multipliers method.

In our application, the classes of the training data set cannot be separated by a linear boundary. For this reason, a polynomial kernel function $\psi(\kb)$ is used to map the input points to a transformed predictor space where a linear boundary can be identified.

After training, new input points are classified according to the sign of a score function, that is the equation of the separating hyperplane; this quantifies the distance of the input from the hyperplane, therefore determining its class:

\begin{equation}
f(\kb) = \sum\limits_i{\sigma_i a_i <\psi(\kb),\psi(\kb_i)>} + b
\end{equation}

\noindent where $\sigma_i$ are the Lagrange multipliers computed during training.
The sign of the score function $\tilde{\phi}_a(\kb) = \textrm{sgn}(f(\kb))$ constitutes our surrogate for the assessment function $\phi_a(\kb)$, suitable to run online and used for the RUL estimation process.

		\subsubsection{Online}
		
The SVM trained offline is employed as a surrogate assessment function to speed up the real-time computations involved in RUL estimation.

The fault propagation rate can usually be described by an Ordinary Differential Equations (ODE) model, accounting, in the most general case, for the current health condition of the system, the environmental and operating conditions, and the expected mission profile.
The evolution of the system health status is computed through the numerical integration of this ODE model. The initial condition is set as the fault vector $\kb$ estimated in the previous FDI phase.
At each integration time step $t_i$ the surrogate assessment function $\tilde{\phi}_a(\kb)$  determines whether the current fault vector $\kb(t_i)$ corresponds to a healthy system or not. 
Since this has to be evaluated iteratively, the use of the full model-based assessment function $\phi_a(\kb)$ would result in long computational times, not suitable for real-time evaluation.
When a faulty condition is detected by the (surrogate) assessment function, the integration is stopped. At this point we can assume:

\begin{equation}
\textrm{RUL} = t
\end{equation}

\noindent where $t$ is the current integration time. That is, the Remaining Useful Life of the system is assumed to be equal to the timestep when the system transitioned from a healthy condition to a faulty one.

\section{Diagnostics and Prognostics of Aircraft Actuation Systems} \label{application}

We develop and demonstrate our methodology for the real-time prediction of Remaining Useful Life for aircraft actuators.
Actuation systems involve the interaction of several, heterogeneous engineering disciplines, such as electronics and software, electrical machines, mechanical systems, hydraulics, structures, thermal dissipation problems, fluid dynamics, vibrations, and tribology.
Additionally, damage propagation may be affected by operating conditions that are not completely predictable, as opposed to similar actuation devices employed for static applications, such as for industrial automation.
As a result, faults affecting such systems have effects on performances that are difficult to predict, and the associated detailed models are computationally expensive. 

Computational methodologies intended to address the open challenges of diagnostic and prognostics for aircraft actuators are of critical interest for mission reliability and cost effectiveness of the whole fleets. Failures in such subsystems can lead to increased down time of the vehicle and may require risk mitigation, since most of these devices are safety critical.

Similar actuation technologies are employed in different fields of engineering, sharing the same open challenges regarding health monitoring and management. As an example, the failure of an actuator on a production line can require the shutdown of the whole production line for repair, with significant income losses.
Given the multidisciplinary nature of the considered application, computational techniques developed to address the prognostic analysis of servo actuators can be extended to deal with health monitoring of similar components not necessarily within the domain of actuation systems. Specifically, any dynamical system involving power electronics, electrical machines, sensors, or mechanical and hydraulic power transmission, can be a potential application of the proposed health monitoring strategy.

	\subsection{Problem Setup} \label{app}	

The particular application addressed in this paper is the real-time estimation of Remaining Useful Life for an Electromechanical Actuator (EMA) for aircraft Flight Control Systems from on-board measurement of the motor current. EMAs
\cite{Sensor2000,Bennouna_2013,Leondes_2014,Mare_2017}
exploit an electric motor coupled to a mechanical transmission to convert power from the aircraft electric system into mechanical power to move the flight control system aerodynamic surfaces. Those actuation systems are commonly employed in small scale UAVs and for secondary flight controls of larger manned aircraft.

An overall weight reduction, compared to the more traditional hydraulic and electrohydraulic systems, can be achieved employing electromechanical actuators to power the whole flight control system of an aircraft (as highlighted by the More Electric Aircraft and All Electric Aircraft design approaches
\cite{Quigley, Howse_2003, Hussien_2012, Deng_2014, Garcia_Garriga_2018}). 
The elimination of a centralized hydraulic power generation system is particularly advantageous for the weight budget of the smaller vehicles, although the power density of an electric actuator is lower than that of an hydraulic one \cite{Weimer,CRONIN_1983,Sharma_2015}. 
Moreover, maintenance on electric systems is easier than on hydraulic ones, since there are no issues related to fluid leakages and contamination.

On the other hand, EMAs are not yet widely employed on safety critical functions for manned aircraft. This is mainly due to the presence of a complex mechanical reducer between the motor and the aerodynamic surface, that introduces the risk of mechanical jamming as a possible failure mode \cite{Stridsberg_2005,Annaz_2008,Missala_2014,Hussain2018}.
This eventuality can lead to the impossibility to control the aircraft with catastrophic consequences.
The introduction of accurate and reliable PHM techniques would increase the safety of operations. Then, a more widespread use of EMAs in larger manned and unmanned vehicles would be allowed, enabling to exploit their advantages on weight and power budget \cite{Zhou_2010,Glennon_1998}.

\begin{figure}[t!]
\centering
\includegraphics[width=.8\textwidth]{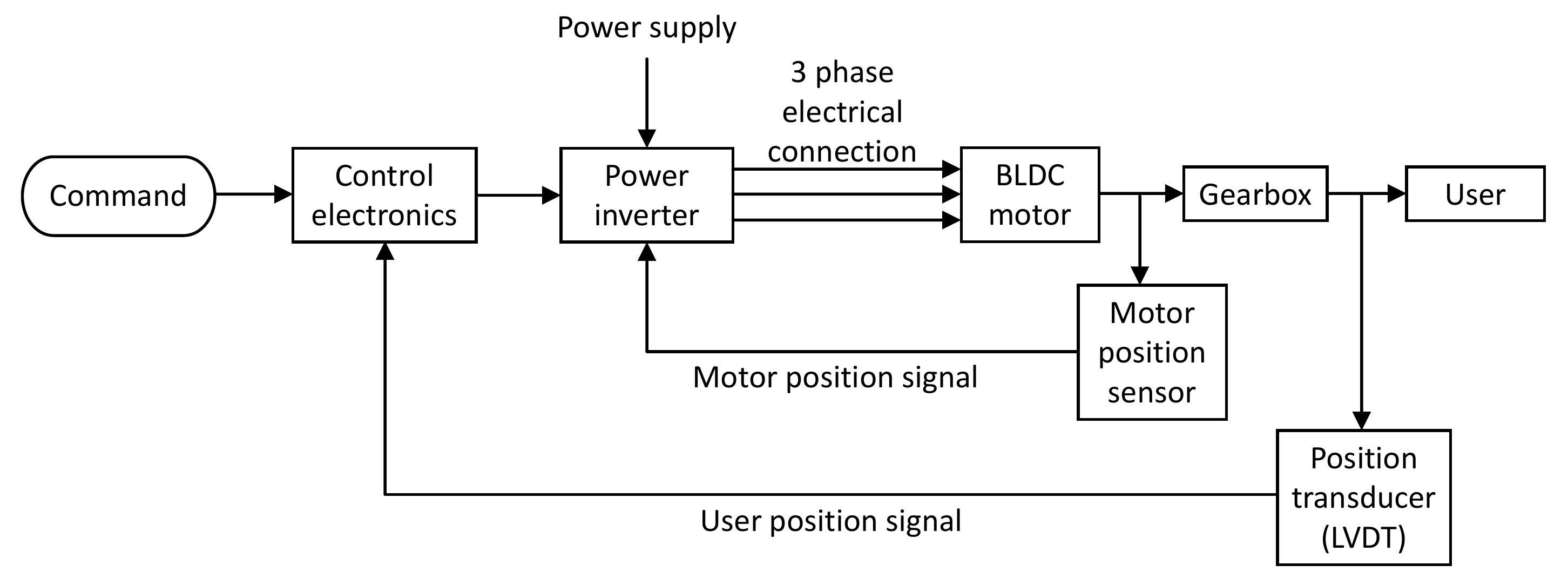} \
\caption{Architecture of the Electromechanical Actuator}
\label{EMAbd}
\end{figure}

The block diagram of Figure \ref{EMAbd} shows the architecture of the considered EMA. This includes a BrushLess Direct Current (BLDC) electric motor along with its power and control electronics, a reducer with a high gear ratio to increase the torque for the user, and a Linear Variable Differential Transformer (LVDT) position sensor to close the feedback loop.

For this study, we consider the effects of five different failure modes, chosen among the most common for EMAs \cite{Balaban_2009,Balaban2009ExperimentalDC,Balaban_2010} and characterized by a slow propagation rate, to allow an effective estimation of the Remaining Useful Life. Those are namely variations in dry friction ($k_1$) and backlash ($k_2$), partial short circuit of each of the three stator phases ($k_{3,4,5}$), rotor static eccentricity ($k_{6,7}$) and controller proportional gain drift ($k_8$); this results in a fault vector $\kb = [k_1, k_2, k_3, k_4, k_5, k_6, k_7, k_8]$ of $n_k=8$ elements. The complete definition of the fault vector $\kb$ is illustrated in Table \ref{tab:Ktable}.
We chose the stator envelope current as the informative variable $y$ to monitor for the prognostic analysis. The reason for this choice is twofold: it is highly sensitive to a number of fault modes and can be easily measured in a physical system; in addition, in many cases stator currents are already measured with the purpose of closing a current feedback loop.

\begin{table}[bp]
  \footnotesize
  \centering
  \caption{Definition of the fault vector $\kb$}
    \begin{tabular}{rrrr}
    \hline
    \textbf{fault parameter} & \multicolumn{1}{c}{\textbf{fault mode}} & \multicolumn{1}{c}{\textbf{lower bound ($k_i = 0$)}} & \multicolumn{1}{c}{\textbf{upper bound ($k_i = 1$)}} \\
    \hline
    \multicolumn{1}{c}{$k_1$} & dry friction & nominal friction & 300\% of nominal friction \\
    \multicolumn{1}{c}{$k_2$} & backlash & nominal backlash & 100 times nominal backlash \\
    \multicolumn{1}{c}{$k_3$} & phase A short circuit & no short circuit & full short circuit \\
    \multicolumn{1}{c}{$k_4$} & phase B short circuit & no short circuit & full short circuit \\
    \multicolumn{1}{c}{$k_5$} & phase C short circuit & no short circuit & full short circuit \\
    \multicolumn{1}{c}{$k_6$} & rotor eccentricity & no eccentricity & eccentricity equal to air gap width \\
    \multicolumn{1}{c}{$k_7$} & eccentricity phase & $-180^o$ & $180^o$ \\
    \multicolumn{1}{c}{$k_8$} & proportional gain drift & 50\% of nominal gain & 150\% of nominal gain \\
    \hline
    \end{tabular}%
  \label{tab:Ktable}%
\end{table}%

	\subsection{Physical models of system dynamics} \label{EMAmodels}

Two models of the actuator with different fidelity are employed: a High Fidelity (HF) model (Section \ref{HFmodel}) is used only offline as the source of reference data: for the high accuracy of the model this dataset can be used as a good emulator of ground truth reference data \cite{Berri_2016}. A Low Fidelity (LF) model (Section \ref{LFmodel}) is used within the assessment function to determine the frequency response of the actuator and compare it with its requirements. The accuracy of the LF model is considered suitable for the sake of this task; the computational cost of the HF model would be impractical for the iterative evaluation required by the assessment function, even for offline execution.
Section \ref{damPropMdl} describes the model for the damage propagation rate, employed for the RUL estimation process.

		\subsubsection{High Fidelity (HF) model} \label{HFmodel}

The HF model is the accurate dynamical model of the EMA, simulating in detail the physical behavior of the actuator subsystems and components. The model accounts for the effects of Pulse Width Modulation (PWM) three-phase current control logic of the motor power electronics, and includes a complete lumped parameters model of the electromagnetic coupling between stator and rotor. This HF model is employed as a simulated test bench, to compute the reference data in replacement of a physical system.

\begin{figure}[t!]
\centering
\includegraphics[width=\textwidth]{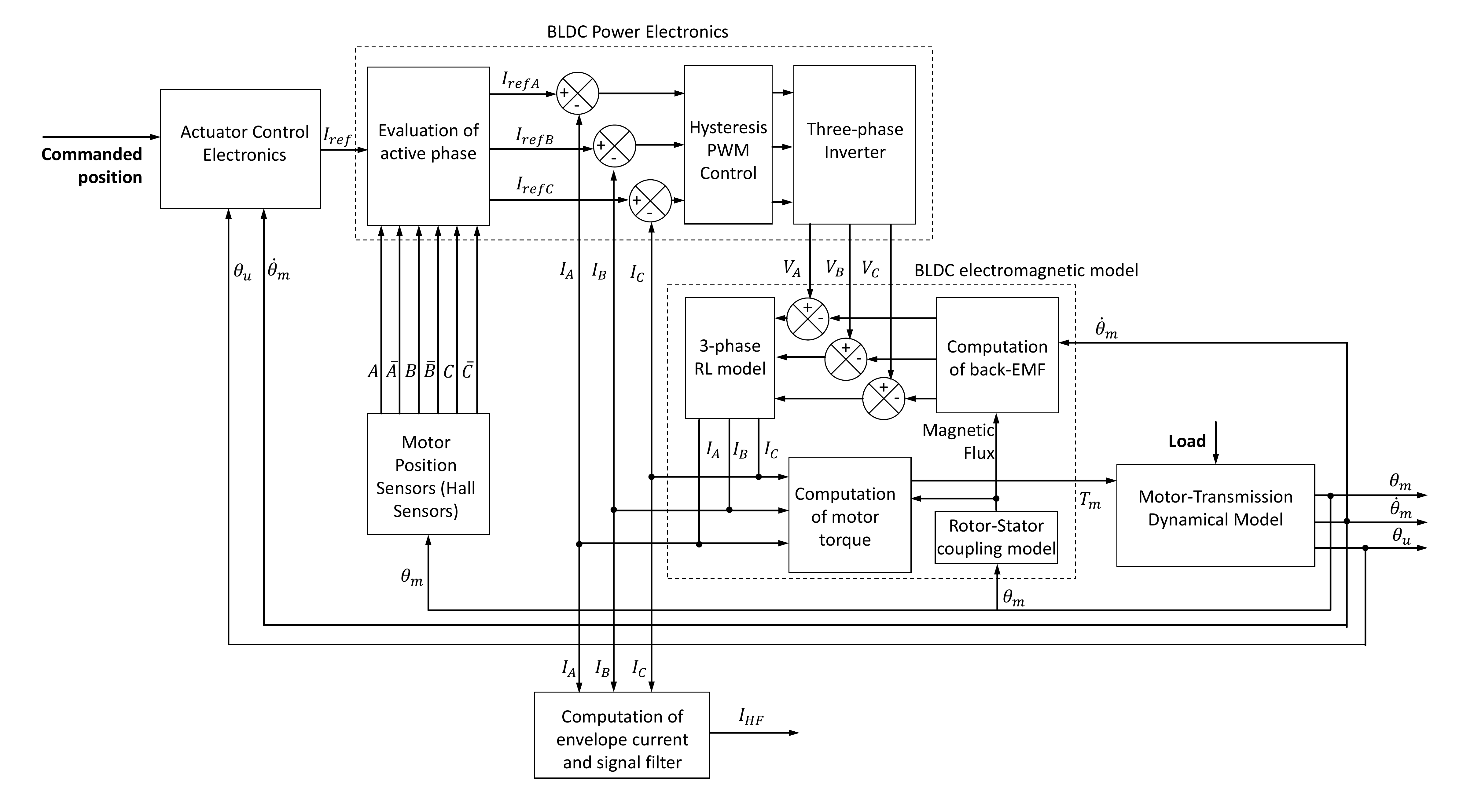} \
\caption{Detailed block diagram of the High Fidelity Model}
\label{figure:HFmodel}
\end{figure}

The architecture of the model is shown in the block diagram of Figure \ref{figure:HFmodel}.
The Actuator Control Electronics block simulates the EMA controller, which compares the commanded and actual positions to compute a reference current signal $I_{r\!e\!f}$ for the motor.
The BLDC Power Electronics subsystem contains the model of the three-phase inverter used to power the BLDC motor. It applies the needed voltages ($V_A$, $V_B$ and $V_C$ for the three motor phases, respectively) on the motor windings to produce to produce the currents $I_A$, $I_B$ and $I_C$ required by the controller; the Hall sensors measure the rotor position $\theta_m$ that is then used to synchronize the phase commutation with the motor rotation.
The BLDC Electromagnetic model contains the rotor-stator coupling model to evaluate the magnetic flux across the air gap. This is employed to compute both the counter-electromotive forces on the stator windings and the torque $T_m$ produced by the motor.
The motor-transmission dynamical model is a second-order model of the actuator mechanics and computes the angular positions $\theta_m$ and $\theta_u$ of motor and user respectively, accounting for several nonlinear effects such as backlash, dry friction and mechanical endstops.
The signal acquisition block computes the envelope $I_{HF}$ of the three phase currents and applies a low-pass signal filter to suppress the high frequency noise produced mainly by the PWM control and obtain the output signal $\yb$ employed for the PHM analysis.

The HF model is implemented in the Matlab-Simulink simulation environment. Its accuracy comes at the expense of a relatively high computational effort. The simulation of a reference 0.5 seconds test signal takes about one minute on a common laptop PC, making this model unsuitable for real-time applications.

		\subsubsection{Low Fidelity (LF) model} \label{LFmodel}

\begin{figure}[b!]
\centering
\includegraphics[width=0.9\textwidth]{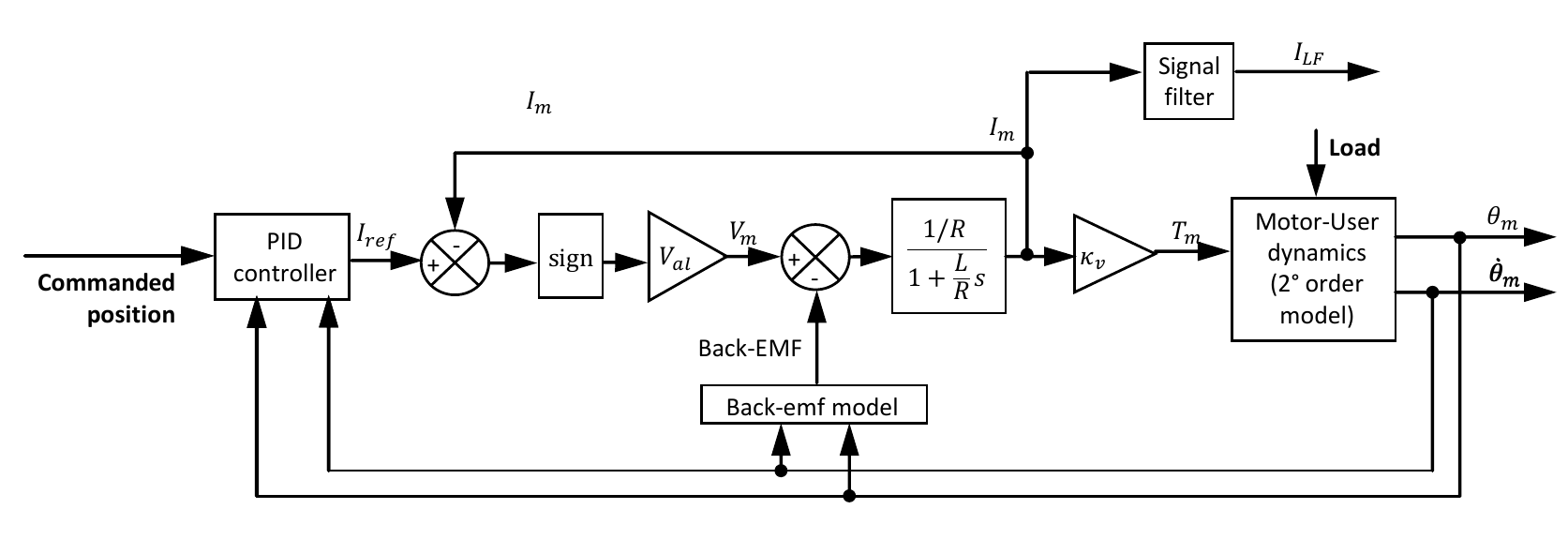} \
\caption{Block Diagram of the Low Fidelity Model}
\label{figure:LFmodel}
\end{figure}

The LF model is a simplified dynamical representation of the same EMA, with complex subsystems represented by simpler blocks. This model is used iteratively to compute the assessment function $\phi_a(\kb)$ (see Sections \ref{RULest} and \ref{damPropMdl}).

The block diagram of the model is shown in Figure \ref{figure:LFmodel}.
The most computationally expensive sections of the HF model are the three-phase inverter and the computation of the magnetic flux across the air gap. Those subsystems are differently handled in the LF model, replaced by a first order DC model whose governing equation directly relates the motor current $I_m$, voltage $V_m$ and torque $T_m$ through the back-Electromotive Force (back-EMF) coefficient $\kappa_v$:

\begin{equation}
RI_m + L\dot{I}_m = V_m - \kappa_v\omega
\end{equation}

\begin{equation}
T_m = \kappa_v I_m
\end{equation}

\noindent where $R$ and $L$ are the stator resistance and inductance, and $\omega$ is the motor angular speed.
This simplified model does not allow accounting directly for the electric faults of partial short circuit and rotor eccentricity. For this reason, Berri et al. \cite{Berri_2016} proposed to model those failure modes by introducing two shape functions to modulate the motor parameters as a function of the angular position of the rotor. The same approach is adopted in this paper: the shape functions are fitted to emulate the waveform of the back-EMF of the HF model in presence of the eccentricity and short circuit fault modes.

The computational time needed for the execution of the LF model is about two orders of magnitude lower than that of the HF model. The average root mean square discrepancy between the HF and LF model is in the order of 1\%.

		\subsubsection{Damage propagation model} \label{damPropMdl}

The model of damage propagation is an ODE based model assuming that each fault grows linearly with the others. It is expressed as:

\begin{equation}
\label{damgrowth:equation}
\dot{\kb}(t) = \textrm{F}\Delta\kb(t) + \boldsymbol{\epsilon}
\end{equation}

\noindent where $\dot{\kb}$ denotes the fault growth rate, $\Delta\kb(t) = \kb(t) - \kb_0$, $\kb_0$ is the fault vector in nominal conditions, $\boldsymbol{\epsilon}$ is an independent identically distributed normal noise, and $\textrm{F}$ is a square matrix whose $F_{ij}$ element expresses the influence of the $j$-th fault parameter on the growth rate of the $i$-th fault parameter.
The matrix $\textrm{F}$ depends on the physics of the system, and can be identified from field data.
The integration is stopped when the assessment function $\phi_a(\kb)$ indicates that the fault condition $\kb$ achieves damage levels that jeopardize the system performance.
The assessment function calls the LF model iteratively with a set of input frequencies to compute the system Bode diagram (it would be very impractical to use the HF model, even for offline computations). Then, the phase margin, gain margin and cutoff frequency are compared to the thresholds imposed by performance requirements to determine whether the actuator is working correctly or not. The computational time of the assessment function on a common laptop PC is in the order of 10 seconds: despite the use of the LF model, the computational cost is not compatible with real-time RUL estimation; this motivates the need of developing a surrogate model for the assessment function to handle the task efficiently (Section \ref{SVMmodel}).

The integration of the ODE is affected by uncertainty due to the noise $\boldsymbol{\epsilon}$ and produces a large dispersion.
In this paper, we handle uncertainty propagation by performing the integration several times for a given starting condition. Then, a gaussian probability distribution is fitted over the RUL estimates, and the values corresponding to 5\%, 50\% and 95\% probability are saved.
In order to obtain a deterministic algorithm, the uncertainty component $\boldsymbol{\epsilon}$ is replaced with a disturbance $\delta$ on the initial fault vector $\kb(t=0)$.
$\delta$ assumes three values computed with a bisection method and calibrated to produce the aforementioned 5\%, 50\% and 95\% probability RUL estimates.

\section{Results and Discussion} \label{app_res}

This section discusses the results obtained with the application of the offline and online phases of the PHM methodology to the specific problem considered in this paper.
A training set is collected from data computed with the physics-based models of Section \ref{EMAmodels}. Specifically, a set of $n_s=10000$ fault combinations is computed with the importance sampling strategy described in Section \ref{MLH}, to obtain the matrix $\textrm{K}$. In our application, $k_7$ is usually excluded by the distribution rescaling process. This parameter encodes the phase of the rotor static eccentricity with respect to the stator windings; then, its probability distribution is necessarily uniform, and should not be modified.
The HF model of the actuator (Section \ref{HFmodel}) is evaluated for each fault combination; a test command is employed, consisting in a linear chirp characterized by a 0.5s duration, $5\cdot10^{-3}$rad amplitude, 0Hz start frequency and 15Hz end frequency.
The output signals $y(t,\kb)$ are acquired with a constant frequency of 20kHz (resulting in $n_e=10001$), necessary to capture the information required by the FDI, and then assembled into the columns of the measurement matrix $\textrm{Y}$.
The assessment function described in Section \ref{damPropMdl} provided the $\textrm{``healthy''}$ or $\textrm{``faulty''}$ labels associated for each fault combination, to be stored into the matrix $\Phi$.
Additionally, two validation sets are assembled to assess the performance of the proposed strategy.
A first validation set is computed to assess the offline training of surrogate models and the online compression and FDI (Sections \ref{offline_res} and \ref{results}). This includes 500 fault combinations, sampled as per Section \ref{MLH}, and the associated signals and values of the assessment function.
A second validation set is employed to assess the online RUL estimation procedure (Section \ref{res3}). This dataset includes 100 fault combinations, the corresponding HF signals and assessment function values, and the RUL (with the associated uncertainty) computed through the damage propagation model described in Section \ref{damPropMdl}. The validation of the online procedure requires the initial faults to be small (i.e. near the nominal condition); otherwise, the system would be already faulty, its remaining Useful Life would be null, and the real time RUL estimation could not be properly tested. Therefore, this second test set is sampled with the procedure of Section \ref{MLH} on a restricted domain in $\kb$, to include mostly $\textrm{``healthy''}$ conditions.

The following sections describe the application of our methodology to the problem of RUL estimation for aerospace EMAs, described in Section \ref{application}.
In particular, Section \ref{offline_res} shows the offline learning of surrogate representations of the models that describe the physics of the considered problem; Section \ref{results} discusses the results of the online procedure.
 All computations were performed on a desktop PC with a i5 3330 quad-core processor @3.00GHz and 8GB of memory, running Windows 10 and Matlab R2016a

	\subsection{Offline: learning models for efficient surrogate representations} \label{SurrogateModels} \label{offline_res}

The following paragraphs discuss the application of the offline procedure to the specific problem addressed in this work: 
the results of the strategy for determining the optimal time coordinates to store and compress the measured signal, derived through the procedure described in Section \ref{compression};
the performance of the model that maps the fault vector $\kb(\boldsymbol{\alpha})$ as a function of the POD coefficients, derived according to Section \ref{FDI-NN};
the outcome of the surrogate model for the assessment function $\phi_a(\kb)$, derived as per Section \ref{SVMmodel}.

		\paragraph{Two-steps data compression} \label{compression_res}

The matrix $\textrm{Y}$ of of the training dataset is employed to learn the informative compression map described in Section \ref{compression}.
$\textrm{Y}$ is used to compute the POD expansion in the form of Equation (\ref{POD:equation1}) whose first $n_m$ modes are retained and used for (i) optimal points selection for signal compression (determined offline) and (ii) POD coefficient reconstruction via gappy POD (to run online).

\begin{figure}[t!]
\centering
\includegraphics[width=.8\textwidth]{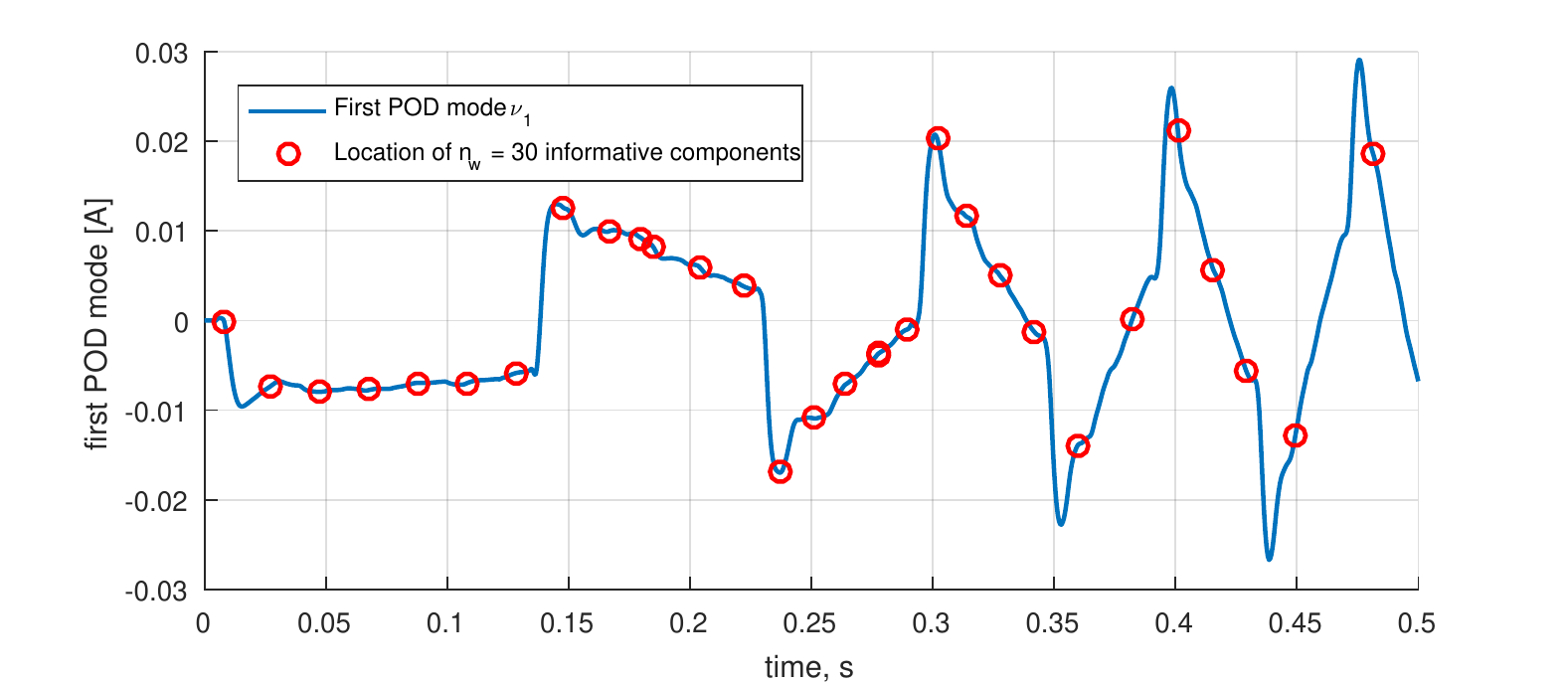} \
\caption{Placement of 30 sampling points over the first POD mode}
\label{res_sampling}
\end{figure}

For the offline identification of the most informative signal points we use the two steps procedure of Section \ref{compression} that allows us to compute a set of $n_w = 30$ points. Those are optimally placed to capture the information of the first $n_m$ POD modes. In this paper, the number of points is chosen to retain sufficient information for the FDI on the base of a previous investigation presented in \cite{BerriMainini18}. Figure \ref{res_sampling} shows an example of placement of the sampling points for $n_m = 1$: the points are not equally spaced in time, but rather tend to be placed by the algorithm in the most significant points to capture the shape of the mode.

				\paragraph{Learning the model for $\kb(\boldsymbol{\alpha})$} \label{FDInetwork}

\begin{figure}[t!]
\centering
\includegraphics[width=.8\textwidth]{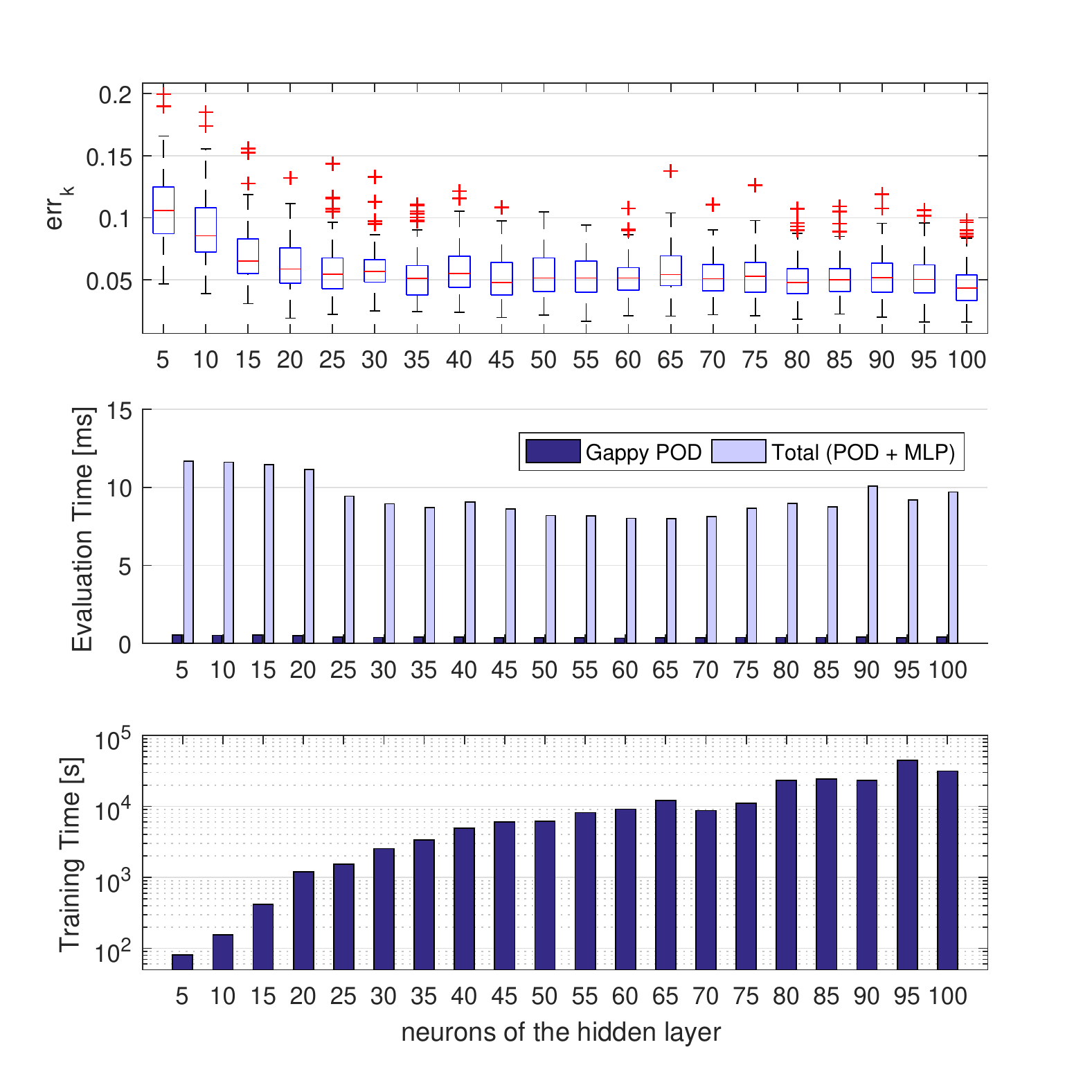} \
\caption{Error in fault identification (top), computational time in evaluation (middle) and in training (bottom) for variable number of neurons in the hidden layer.}
\label{neurons}
\end{figure}

For the FDI step, a Multi-Layer Perceptron with one hidden layer is used to compute the estimated fault vector $\kb$ from the coefficients $\boldsymbol{\alpha}$ reconstructed via Gappy POD.
The training set for the neural network model is composed by the POD coefficients computed in the previous step, and the matrix $\textrm{K}$.
The choice of a suitable number of neurons for the network emerges from a tradeoff between accuracy and computational time.
A study was performed on the number of neurons, by varying the neurons of the hidden layer from 5 to 100, while the number of neurons of the output layer remains constant at $n_k = 8$.
Figure \ref{neurons} shows the computational time of the network in training and in evaluation, as well as the mean squared error in the fault identification plotted against the number of neurons in the hidden layer:

\begin{equation}
err_k = \sqrt{\frac{1}{8}\sum\limits_{i=1}^8 {w_i(k_i^\textrm{estimated} - k_i^\textrm{actual})}}
\end{equation}

\noindent where the weights $w_i$ are all unitary except for $w_7 = k_6^\textrm{actual}$.
The elements $k_6$ and $k_7$ of the fault vector encode the amplitude and phase of the rotor static eccentricity, respectively (see Table \ref{tab:Ktable}). The error on eccentricity phase is then weighted by the actual eccentricity amplitude: this permits to achieve a more physically significant error estimate when the eccentricity is small in amplitude.

By increasing the number of neurons in the hidden layer, the accuracy initially increases and the mean squared error decreases down to 5\% for 20 neurons. Adding more neurons does not produce significant benefits, neither on accuracy nor on computational time in evaluation; conversely, the increased complexity of the model reflects in longer computational times for training. Therefore, we consider the network with 20 neurons in the hidden layer as the most efficient candidate to perform the FDI for the addressed application.
The computational time required by Gappy POD for the estimation of POD coefficients is at least one order of magnitude shorter than that required by the MLP in evaluation; so, the use of POD coefficients instead of the signal does not carry a significant penalty in computational time for real-time FDI.

The fault vector $k^\textrm{estimated}$ computed by the network is employed as initial condition for the RUL estimation.

				\paragraph{Learning the model for the assessment function} \label{SurrAssFcn}

For the surrogate modeling of the assessment function $\phi_a(\kb)$, we employ a polynomial kernel SVM, trained with the matrix $\textrm{K}$ as the input and $\Phi$ as the target.
The SVM was assessed with the 500 signals validation set, and achieved a 98.2\% success rate in emulating the assessment function (see Table \ref{tab:SVM}), with an average computational time in the order of 1ms, thus reducing the computational effort of almost four orders of magnitude with respect to the evaluation time required by the complete assessment function.

\begin{table}[b!]
  \centering
  \caption{Performance of the SVM; test set composed of 500 fault combinations, of which 123 corresponding to a healthy actuator and 377 corresponding to a system failure}
    \begin{tabular}{rrr}
\hline
    correctly detected & 119   & 96,75\% \\

    correctly undetected & 372   & 98,67\% \\
    missed detections & 4     & 3,25\% \\
    false positives & 5     & 1,33\% \\
\hline
    \textbf{total correct} & \textbf{491}   & \textbf{98,20\%} \\
    \textbf{total wrong} & \textbf{9}     & \textbf{1,80\%} \\
\hline
    \end{tabular}%
  \label{tab:SVM}%
\end{table}%

	\subsection{Online: real-time diagnostics and prognostics} \label{results} 

This section presents the application of the online process to the considered problem.
Specifically, to assess the performance of the prognostic framework, first we test the signal compression, the FDI phase, and the RUL estimation phase separately; then, the RUL estimation is employed in combination with the FDI strategy to establish the overall accuracy of the method (Section \ref{res3}).

				\paragraph{Signal acquisition and compression} \label{gappy}

\begin{figure}[b!]
\centering
\includegraphics[trim=0 0.25cm 0 0.1cm, clip, width=1\textwidth]{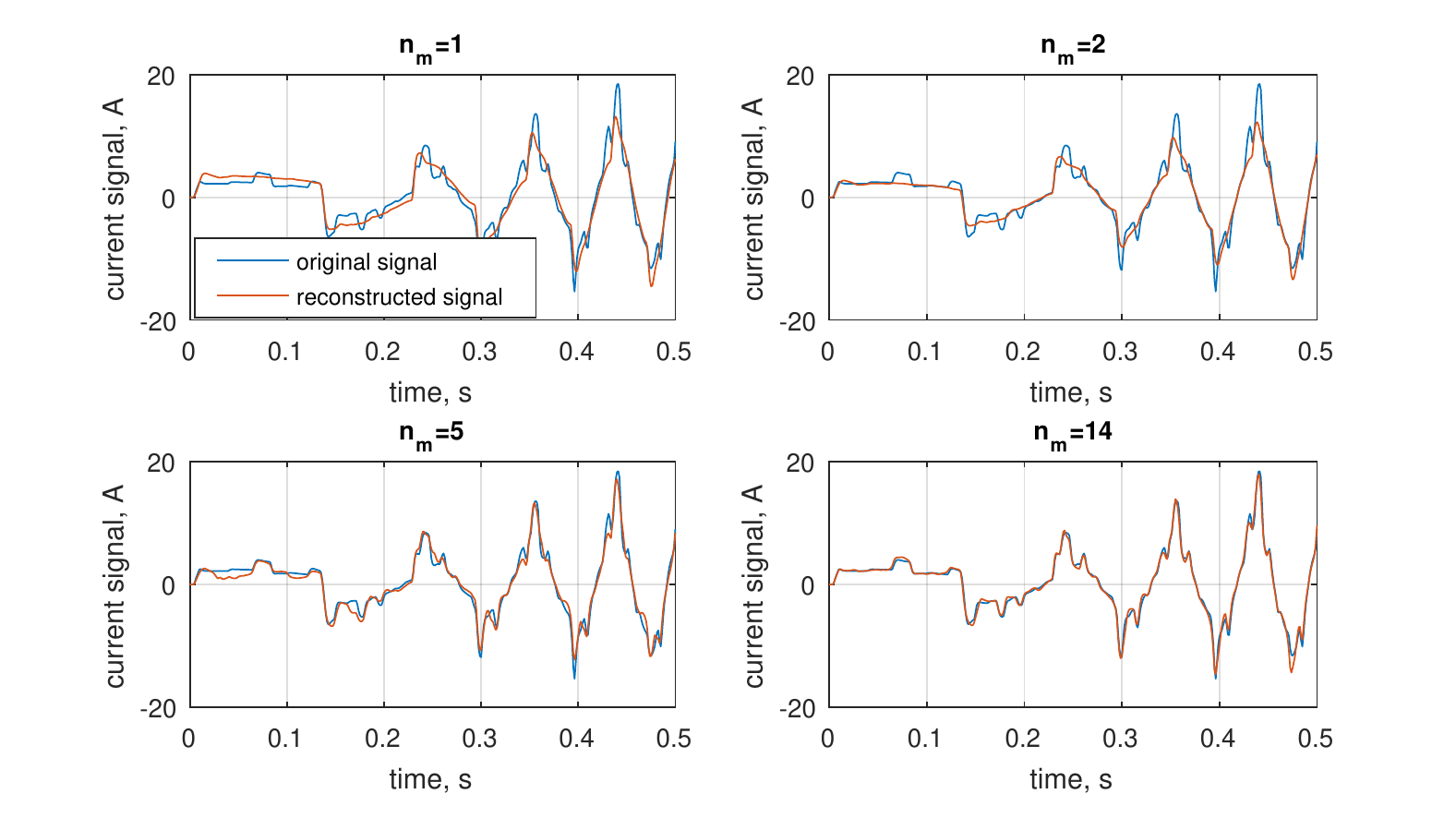} \
\caption{Signal reconstruction via Gappy POD with a variable number of retained modes $n_m$}
\label{rec1}
\end{figure}

The signal is compressed online by storing and processing only the measures corresponding to the time coordinates determined offline through the procedure described in Section \ref{compression}.
Online, Gappy POD is employed to determine the coefficients $\ab(\kb)$ from the stored signal.
Figure \ref{rec1} shows the reconstruction of a compressed signal from the test set via Gappy POD, with an increasing number of POD modes.
To assess the accuracy of the signal estimate, we evaluate the normalized root mean squared error $err_\alpha$ of the POD coefficients $\alpha^{\textrm{gappy}}$ estimated with Gappy POD with respect to the coefficients $\alpha^{\textrm{full}}$ computed using the full signal:

\begin{equation}
err_\alpha = \frac{\sqrt{\frac{1}{n_m}\sum\limits_{i=1}^{n_m} {(\alpha_i^\textrm{gappy} - \alpha_i^\textrm{full})}}}
             {\max{(\ab^\textrm{full})}-\min{(\ab^\textrm{full})}}
\end{equation}

\begin{figure}[t!]
\centering
\includegraphics[width=\textwidth]{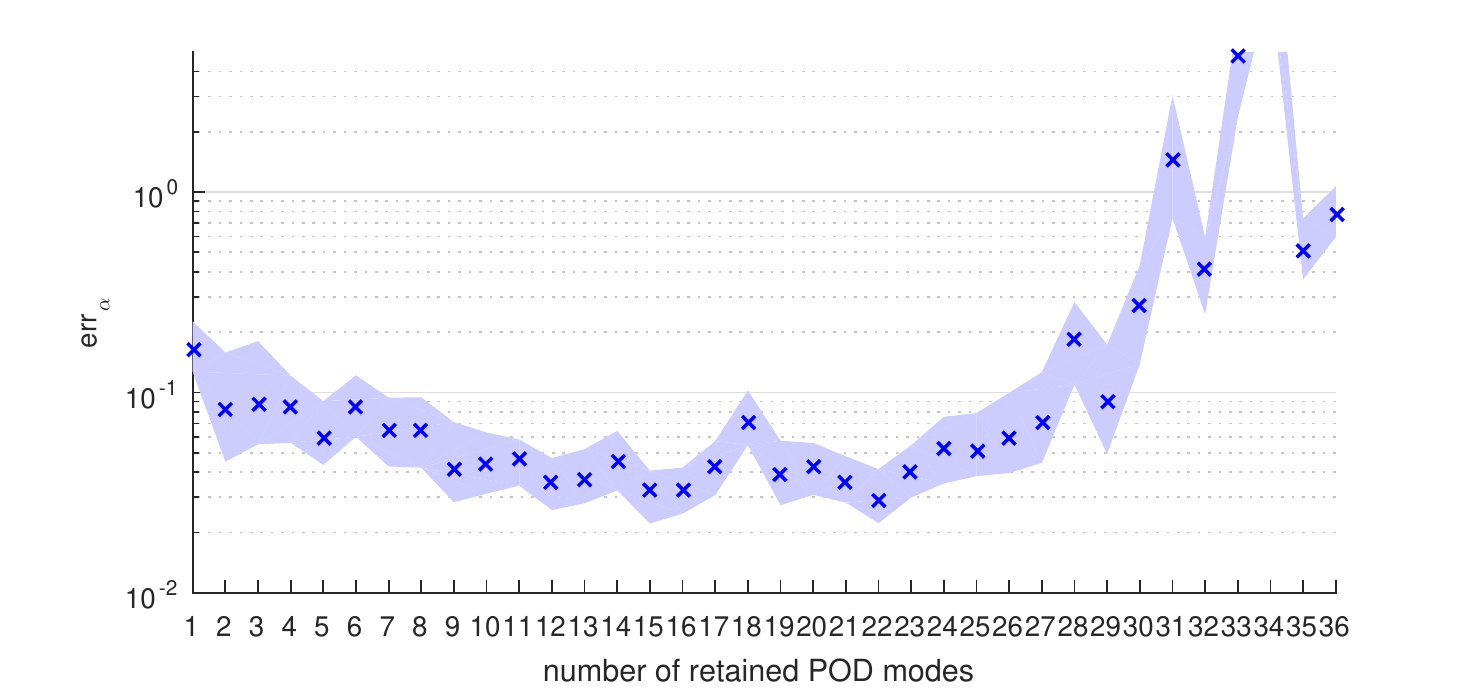} \
\caption{Offline preliminary assessment of Gappy POD coefficients identification, for varying number of modes retained. The blue crosses are the median error over the test set, while the amplitude of the tolerance band is set equal to the interquartile range.
$n_m = 22$ modes result in the best accuracy, with a median error below 2\%
}
\label{res_gappy}
\end{figure}

Figure \ref{res_gappy} shows the error computed for increasing size of the gappy matrix G, that is, for increasing number of dominant components used for signal reconstruction through Gappy POD (according to Equation (\ref{POD:equation1})).

For a number of modes larger than 10, it is already possible to identify the coefficients of the POD modes with an error in the order of 1\%, comparable to the discrepancy yielding for the LF model with respect to the HF model (See Section \ref{EMAmodels}).
We choose to employ the first 22 POD modes for the following steps, corresponding to about 97\% of the snapshot information; as shown in Figure \ref{res_gappy}, for our application this number of modes yields the minimum mean value of $err_\alpha$.
The use of a larger number of modes might increase the risk of including misleading information, affecting the accuracy of the signal reconstruction via Gappy POD. Additionally, with more than 30 modes (i.e. more modes than sampling points) the gappy matrix G becomes ill conditioned; the signal reconstruction is strongly affected by numerical noise with an increase of computational time and an error the same order of magnitude of the signal. A similar behavior has been already observed in literature \cite{Data2decisions}.

				\paragraph{Fault Detection and Identification} \label{res1}

\begin{figure}[b!]
\centering
\includegraphics[width=1\textwidth]{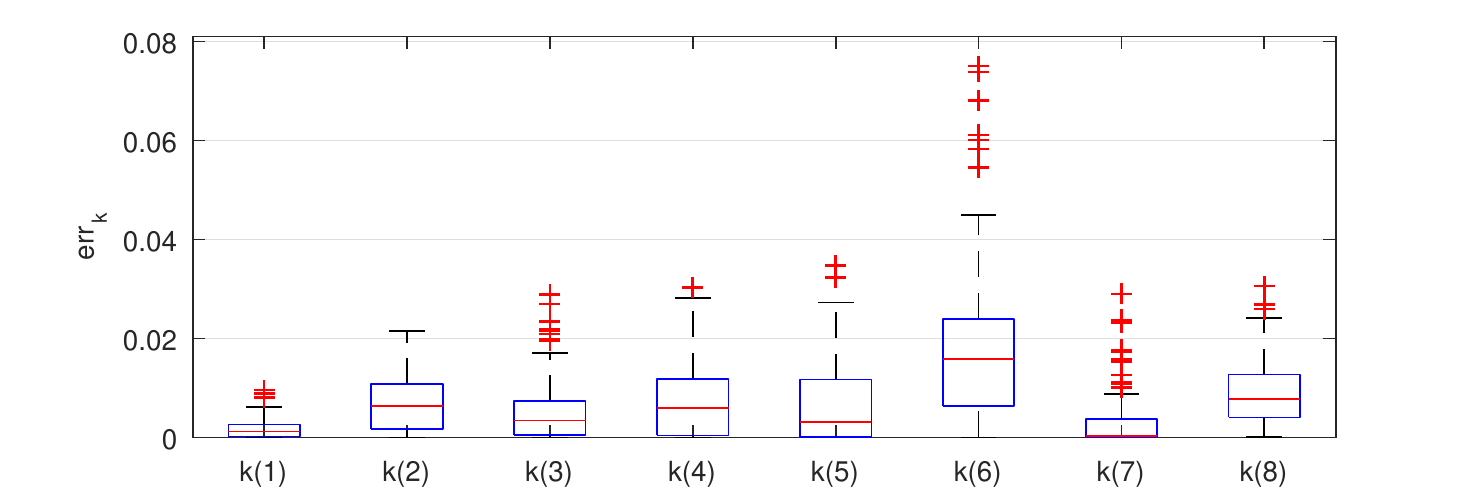} \
\caption{FDI error on each fault parameter}
\label{res_FDI}
\end{figure}

This section describes the assessment of the online performance of the two-layer perceptron for the FDI algorithm.
Figure \ref{res_FDI} shows the error on parameter identification highlighting the contribution of each component of the fault vector.
The average error for each variable $\vert k_i^\textrm{estimated} - k_i^\textrm{actual} \vert$ is at most in the order of 1\%, which is comparable to the discrepancy between the HF and LF model. Then, our data driven FDI technique performs comparably to a more traditional model-based technique, which commonly exploits an online optimization algorithm to match the HF and LF responses \cite{Berri_2017, ESREL2018}.
The FDI accuracy with respect to the individual fault parameters depends on the particular application and on the sensitivity of the monitored variable to the different fault modes. Figure \ref{res_FDI} highlights how, for the considered application, the FDI performs better in the identification of the dry friction fault ($k_1$): the envelope current (i.e. the analyzed output signal $y$) is highly sensitive to this failure mode, resulting in better accuracy. On the other hand, the rotor eccentricity ($k_6$) is identified with a higher error and greater dispersion. This failure mode results in high frequency disturbances of the monitored signal, and the effect of information loss due to the truncation of the POD expansion is worse.
The computational time is in the order of a few milliseconds, allows computations to be run onboard, and enables real-time fault detection.

				\paragraph{Estimate the Remaining Useful Life} \label{res2}
To assess the accuracy of the online RUL estimation procedure alone, we test the performance of our online RUL step leaving out the error contribution associated with the FDI step. To do so, we test the SVM based algorithm using the reference fault vector $\kb^\textrm{actual}$ as initial condition, in place of the estimate computed in the FDI phase $\kb^\textrm{estimated}$.

\begin{figure}[b!]
\centering
\includegraphics[width=.8\textwidth]{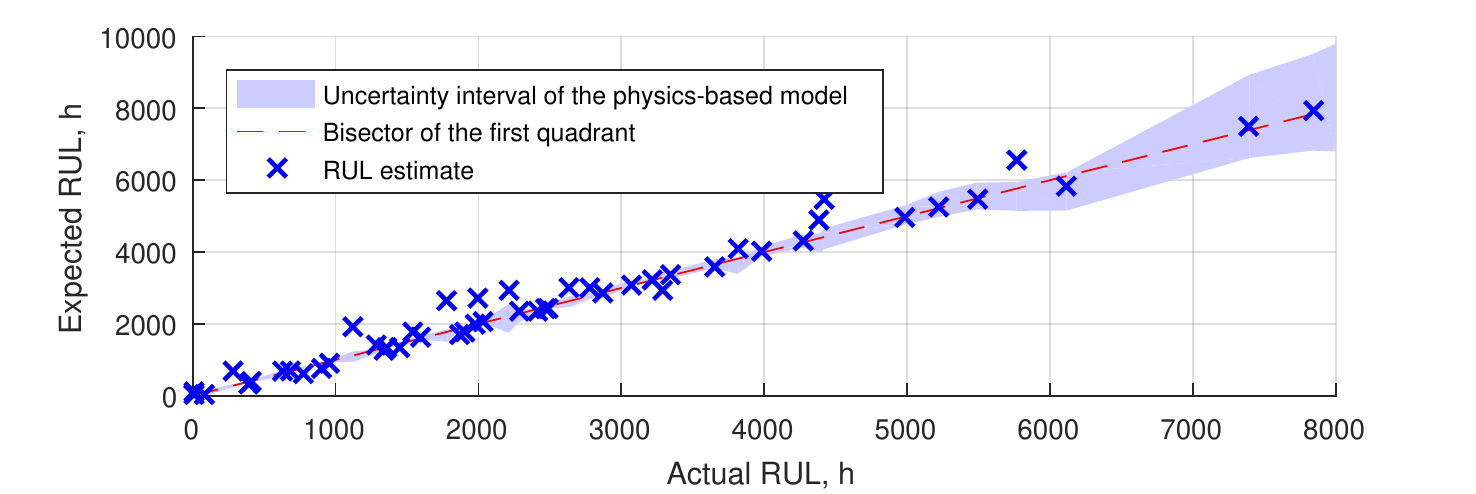} \
\caption{Expected Remaining Useful Life with physics-based assessment function and SVM, assuming the FDI error to be null (i.e. starting from the actual fault condition).}
\label{res_RUL1}
\end{figure}

Figure \ref{res_RUL1} shows the upper and lower bound of the RUL (with 90\% confidence) as a function of the expected RUL.
On the horizontal axis is reported the RUL computed by the full model (i.e. the damage propagation model of Section \ref{damPropMdl} in combination with the assessment function $\phi_a(\kb)$) at 50\% probability, which we assume to be the actual value. The red dashed line is the bisector of the first quadrant, and represents an ideal RUL estimate (i.e. not affected by any error).
The tolerance band represents the uncertainty interval of the full model: its lower and upper bounds correspond to the RUL computed by the full model respectively at 5\% and 95\% probability, respectively.
The blue crosses are the RUL values estimated with the SVM model from the actual initial condition; in most cases, the estimated values fall within the uncertainty interval associated to the reference physics-based model.

	\subsection{Results for the entire real-time PHM information flow} \label{res3}

\begin{figure}[t!]
\centering
\includegraphics[width=.8\textwidth]{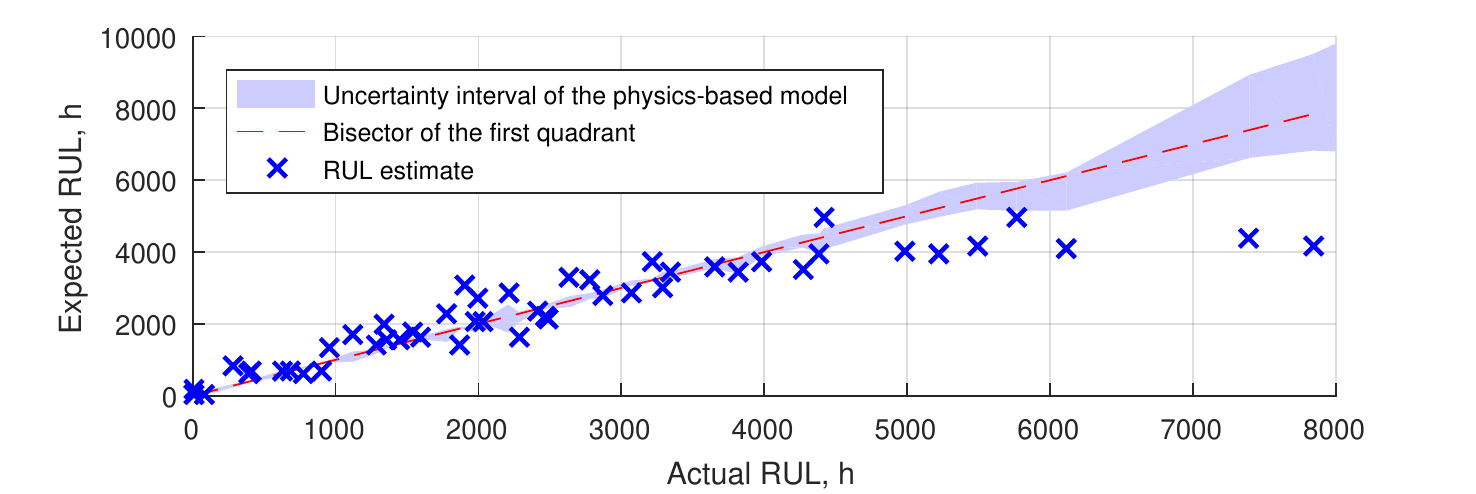} \
\caption{Predicted Remaining Useful Life with estimated fault condition and SVM surrogate assessment function}
\label{res_RUL2}
\end{figure}

A final test assesses the whole real-time PHM flow, including the signal compression and FDI strategy in combination with the RUL estimation method.
The SVM based RUL estimation is computed with the initial condition $\kb^\textrm{estimated}$ estimated through the FDI algorithm described in Section \ref{FDI}; the results, shown in Figure \ref{res_RUL2}, are compared to the expected RUL (with the associated uncertainty interval) computed through the full damage propagation model.
The global RUL estimate is affected by the errors introduced by both the RUL estimation itself and the FDI; then, the uncertainty is necessarily higher than the previous case, resulting in a slightly greater dispersion.
RUL values larger than 4000 hours are commonly underestimated. In these cases, the initial fault parameters assume very small values, that are of the same order of magnitude of the uncertainty associated to the FDI process. The identified fault condition is commonly worse than the actual one, which results in estimating a faster fault propagation and a shorter RUL.
Larger system faults can be detected with higher relative accuracy and the dispersion associated with the RUL estimate decreases.
Therefore, our strategy achieves a higher accuracy when it is needed, that is when 
a failure is about to occur.
When the Remaining Useful Life is long, there is no stringent need to know its value with high precision because there is long time ahead to plan the maintenance strategy at best, purchase spares and schedule replacements.
Additionally, in the first part of the system operational life, an underestimation of the RUL is preferable to an overestimation, for obvious safety reasons, and does not trigger an unnecessary maintenance intervention, since the estimated time to failure is still long.

\begin{figure}[b!]
\centering
\includegraphics[trim=1.5cm 0 1.3cm 0,clip,width=1\textwidth]{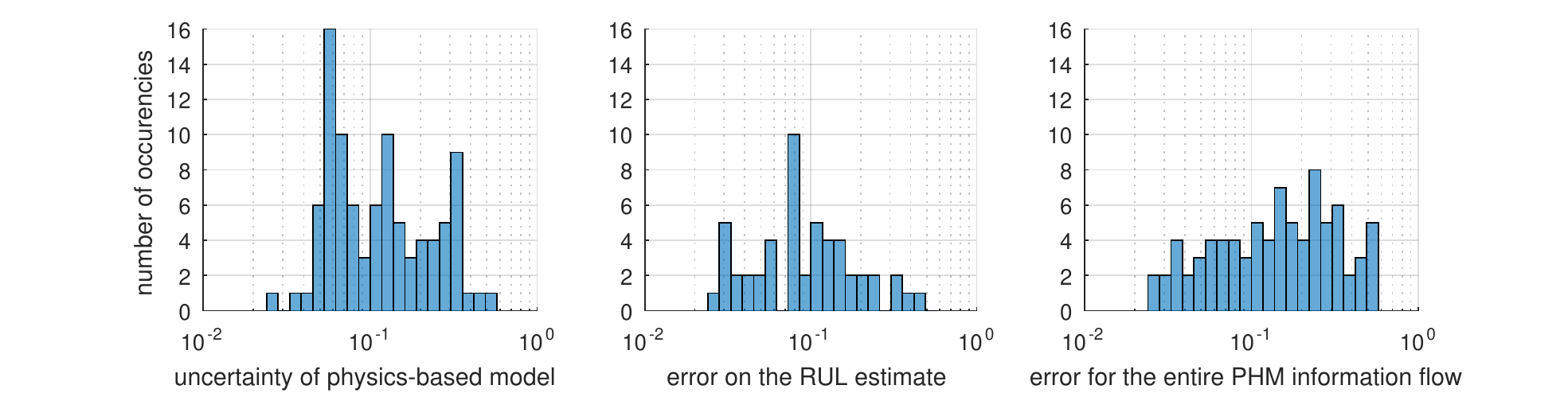} \
\caption{Comparison between the uncertainty of the physics-based model for RUL evaluation (left), the error of the real-time RUL estimation process (middle) and the error of the entire PHM online information flow (right).}
\label{res_overall}
\end{figure}

Figure \ref{res_overall} compares the uncertainty associated to the physics-based model for RUL with the relative error in RUL estimation $err_\textrm{RUL}$ resulting from the RUL estimation alone (starting from $\kb^\textrm{actual}$) and in combination with the FDI (starting from $\kb^\textrm{estimated}$). The error is defined as follows:

\begin{equation}
err_\textrm{RUL} = \frac{\vert RUL_\textrm{estimated} - RUL_\textrm{actual} \vert}{RUL_\textrm{actual}}
\end{equation}

\noindent and the uncertainty of the physics-based model is:

\begin{equation}
\Delta_\textrm{RUL} = \frac{\vert RUL_{95\%} - RUL_{5\%} \vert}{RUL_\textrm{actual}}
\end{equation}

\noindent where  $RUL_{95\%}$ and $RUL_{5\%}$ are the upper and lower bounds of the RUL with 90\% confidence.
The error of the RUL estimation is the same order of magnitude of the uncertainty interval of the physics based model. The error associated with the whole PHM process is still the same order of magnitude, but its distribution is more skewed to the right compared to the error of the RUL estimation alone. This reflects in a higher median error, confirming the underestimation of very long RULs that emerges from the comparison from Figures \ref{res_RUL1} and \ref{res_RUL2}.

The absolute variance of the error is relatively high, being usually in the order of 20\% of RUL, and in some cases almost comparable to the RUL itself.
This behavior is partly due to the inherent uncertainty in the rate of damage propagation, which depends on a multitude of uncontrollable and unpredictable variables, such as environmental conditions and the particular time-history experienced during the system operational life.
However, the error decreases as the damage grows, so the information about the system life is provided with higher accuracy when needed more (when a maintenance intervention needs to be planned). Additionally, this data driven strategy allows for a higher accuracy than the traditional \textit{a priori} estimate \cite{electronicsReliability, mechanicsReliability, Garmendia2015, Venkataraman2017}, and adds virtually no cost for the implementation, since no dedicated hardware is required.
The computational time required online by the whole PHM process is in the order of milliseconds and allows the FDI and RUL estimation to be performed in real-time.
In contrast, traditional model based approaches require computational times that range from minutes to tenths of minutes, which would be completely unsuited for on-board applications.
These strategies \cite{BerriGA,ESREL2019Re} imply solving numerically systems of ODEs with very small integration timesteps, and the indentification is performed over the full dimensional dataset with empirical or semi-empirical optimization algorithms.

\section{Concluding remarks}

A comprehensive methodology for real-time fault detection and prognostics of dynamical assemblies has been proposed.
Our methodological framework leverages a combination of projection-based model reduction and machine learning strategies to achieve reliable and timely estimates of the system useful life.
The method has been developed and assessed for the overall Prognostics and Health Management (PHM) process applied to an electromechanical actuator for aircraft flight control systems.
In addition, a simple model for estimating the fault propagation rate has been proposed, and a custom sampling technique has been employed to capture sufficient information from the system with a limited number of samples, which resulted particularly effective for our specific application.

The strategy permits to achieve an accuracy in FDI and RUL estimation that is comparable to computationally-intensive physics-based methods; at the same time, it requires few online data storage and processing resources, allowing for a fast and reliable on-board, real-time execution: the online computational time for fault detection is reduced by several orders of magnitude with respect to standard computationally expensive, physics-based methods.
The availability of RUL estimate in real-time would permit to efficiently inform adaptive maintenance planning, allowing for significant cost reduction with respect to the standard periodical inspections and replacements, based on the analysis of the average failure rate of the components.
The results show that our strategy for the real-time estimate of Remaining Useful Life allows to achieve high prediction accuracy when the monitored components are approaching the end of their operative life: this permits a dynamic and informed scheduling of maintenance interventions and an adaptive delivery of supplies and spares; additionally, the mission can be dynamically reconfigured to avoid overstressing faulty subsystems.
When the system behaves nominally, our strategy tends to underestimate its useful life; however, this occurrence is safe and does not result in planning the unnecessary replacement of healthy components, since both the actual and estimated RUL are long.

Future developments include the assessment of alternative machine learning strategies for the FDI and RUL estimation, a more exhaustive study on uncertainty propagation, and the experimental validation of the models and algorithms.

\section*{Acknowledgments}
The authors thank Prof. Paolo Maggiore at Politecnico di Torino for his support.
Additional acknowledgments to the Visiting
Professor Program of Politecnico di Torino for the support to Dr. Laura Mainini.

\section*{References}

\bibliography{mybibfile}

\end{document}